\documentclass[letterpaper,notoc]{JHEP3}
\usepackage{psfrag,amsmath,amsthm,amssymb,subfigure,array,verbatim}
\usepackage{url}

\usepackage{psfrag}

\ifpdf     
    \usepackage[pdftex]{graphicx}
    \usepackage{epstopdf}  
\else  
    \usepackage[dvips]{graphicx}
\fi

\newcommand{\be}{\begin{equation}}
\newcommand{\ee}{\end{equation}}
\newcommand{\bea}{\begin{eqnarray}}
\newcommand{\eea}{\end{eqnarray}}

\newcommand{\ra}[1]{ \stackrel{\scriptscriptstyle #1}{ \rightarrow  }}  

\newcommand{\eps}{\varepsilon}

\newcommand{\qbar}{\bar q}

\title{Pure Samples of Quark and Gluon Jets at the LHC}
\author{Jason Gallicchio and Matthew D. Schwartz\\
Jefferson Physical Laboratory, Harvard University, Cambridge, MA 02138}
\date{\today}

\abstract{
Having pure samples of quark and
gluon jets would greatly facilitate the study of jet
properties and substructure, with many potential standard model
and new physics applications. To this end, we consider multijet
and jets+$X$ samples, to determine the purity that can be
achieved by simple kinematic cuts leaving reasonable production
cross sections. We find, for example, that at the 7\,TeV LHC,
the $pp \to \gamma$+2jets sample can provide 98\% pure quark
jets with 200\,GeV of transverse momentum and a cross section of
5\,pb. To get 10\,pb of 200\,GeV jets with 90\% gluon purity,
the $pp \to $ 3jets sample can be used. $b$+2jets is also useful
for gluons, but only if the $b$-tagging is very efficient.
}

\begin{document}


\section{Introduction}

Proton colliders, like the Large Hadron Collider at CERN,
produce an enormous number of high energy jets. These jets are
manifestations of hard quarks or gluons produced at very short
distances, which shower and fragment into collections of
collinear particles. Being able to distinguish quark and gluon
jets could be extremely useful for new physics searches.
For example, many models with
supersymmetry produce dominantly quark jets while
their backgrounds are dominantly gluon jets.
The hope is then to discriminate signal from
background by using observables like jet mass,
which are strongly correlated with
flavor~\cite{Kaplan:2008ie,Han:2009em,Thaler:2010tr,Gallicchio:2010dq,Cui:2010km,Abdesselam:2010pt,Butterworth:2008iy,Krohn:2009wm}.
In order to validate these observables on data,
it would be useful to have relatively pure samples of light quark or gluon jets
to study. It is the purpose of this paper to suggest where those samples might be found.
%

At leading order in perturbation theory, there is no ambiguity in what is meant by the quark
and gluon jet fraction in any exclusive sample.
For example, as we show below, in a 300\,GeV dijet sample at the
7\,TeV LHC, the division is roughly 50/50. This comes simply from the ratio of
the LO cross sections for the various channels, which do not interfere.
The fraction can be defined beyond leading-order as well. In fact,
it is well-defined to to all orders in perturbation theory up to the same power corrections
that affect any jet algorithm's parton correspondence.
These power corrections involve the jet size $R$ (equivalently
the jet's mass-to-energy ratio $m/E$) and $\Lambda_{\mathrm{QCD}}/E$.
One can also define an infrared-safe definition of flavor at the jet level~\cite{Banfi:2006hf}, but that is not the subject of this paper.
We further discuss the theoretical issues associated with defining quark and gluon jets in Section~\ref{sec:theory}.

To be clear, we do not propose that the quark and gluon fractions
can be measured directly in data.
Instead, one can measure observable properties of the samples, such as the jet mass, and compare them to theoretical
predictions, such as from Monte Carlo simulations. The purity calculations in this paper suggest regions where the measurements would be most enlightening.

It may not be obvious why one would want pure samples of quark or gluon jets at all.
Instead, one could just study the jet
observables directly in any mixed sample.
For example, it is well known that the distribution
of jet mass for 300\,GeV jets is typically wider and peaks at
larger values for gluon jets than quark jets.
In a 50/50 sample, such as the 300\,GeV dijet sample,
one could then to hope to find two separated peaks.
Unfortunately, the combined distribution does \emph{not} have
two distinct peaks for jet mass, or charged particle count, or
any other known discriminant --- the distributions are just
too broad. Moreover,
correlations in the 2D distribution of observables like jet mass and
charged particle count might take different forms that would be
impossible to see in a 50/50 sample.
The purer the sample, the closer
one can come to studying quark and gluon jets on an
event-by-event basis.

In this paper, we simulate a wide variety of processes at tree
level for the 7\,TeV LHC. These include events with gluon and
light quark ($uds$) jets, $b$-jets, $W$'s, $Z$'s and $\gamma$'s.
We begin using only
the experimentally minimal cuts. Then we find kinematic cuts,
such as on rapidity differences, which further purify the samples. Section~\ref{sec:start} describes
the event samples and Section~\ref{sec:pure} the purification procedure.
Section~\ref{sec:theory} discussed theoretical issues associated with defining quark and gluon fractions in perturbation theory.
Section~\ref{sec:conclusion} summarizes the results.


\section{Starting Samples to Explore and Purify \label{sec:start}}

All events were generated with {\sc madgraph
v4.4.26}~\cite{Alwall:2007st}, a tree-level matrix element
generator, using leading order {\sc CTEQ6L1}
PDFs~\cite{Stump:2003yu}. Working only at tree-level makes our
results independent of any jet-algorithm and
showering/hadronization routine. Of course, we do not expect
the efficiencies we find to agree with efficiencies one would
get after full simulation, or in data, but this is a simple and
informative way to determine where quark and gluon jets can be
found.

For each sample and each $p_T$, 200,000 events were
generated with the following cuts:
\begin{itemize}
    \item $p^j_T > p_T$ for all `jets', meaning any gluons or $uds$ quarks.
    \item $p^\gamma_T > 20$\,GeV for any photons
    \item $p^\ell_T > 20$\,GeV for any leptons from $W$ or
        $Z$ decays (including missing $E_T$ from neutrinos)
    \item $p^b_T > 20$\,GeV for any $b$ quarks.
    \item $|\eta|<2.5$ for any jet, $b$, photon, or charged lepton.
    \item $\Delta R>1.0$ between any two jets.
    \item $\Delta R>0.5$ between any jet and any photon or between any jet and charged lepton.
\end{itemize}

Since the quark and gluon fractions, as well as jet
properties, can be strongly dependent on $p_T$, we
have to be careful about how we divide the sample into
different $p_T$ bins. We will often find that it is the softest
jet in a sample, such as the softest jet in the 3jet or
$\gamma$+2jet sample, which leads to the highest purity. Since
the cross sections fall rapidly with $p_T$, the majority of
events for a given $p_T$ cut will fall around that minimum
$p_T$. This is why \emph{all} jets in a given
sample must be above the given $p_T$, with `jet' here referring only
to light quarks or gluons. In the 200\,GeV $bjj$ sample, for
example, each light quark or gluon is required to have a $p_T
\ge 200$\,GeV, but the $b$ is only required to have a $p_T \ge
20$\,GeV. In 2-object final states like $\gamma$+jet, the
$\gamma$ automatically also satisfies the jet $p_T$
requirement.

Samples where only one jet satisfies the
hard $p_T$ cut, with the others having a $p_T > 20$\,GeV cut,
were also examined. These have larger cross sections,
but only the hardest jet tends to fall within
the $p_T$ range of interest, and the kinematic cuts required to achieve
high purities reduced the cross section below the softest-jet samples
discussed here.

\begin{figure}[h]
\begin{center}
\includegraphics[width=0.90\textwidth]{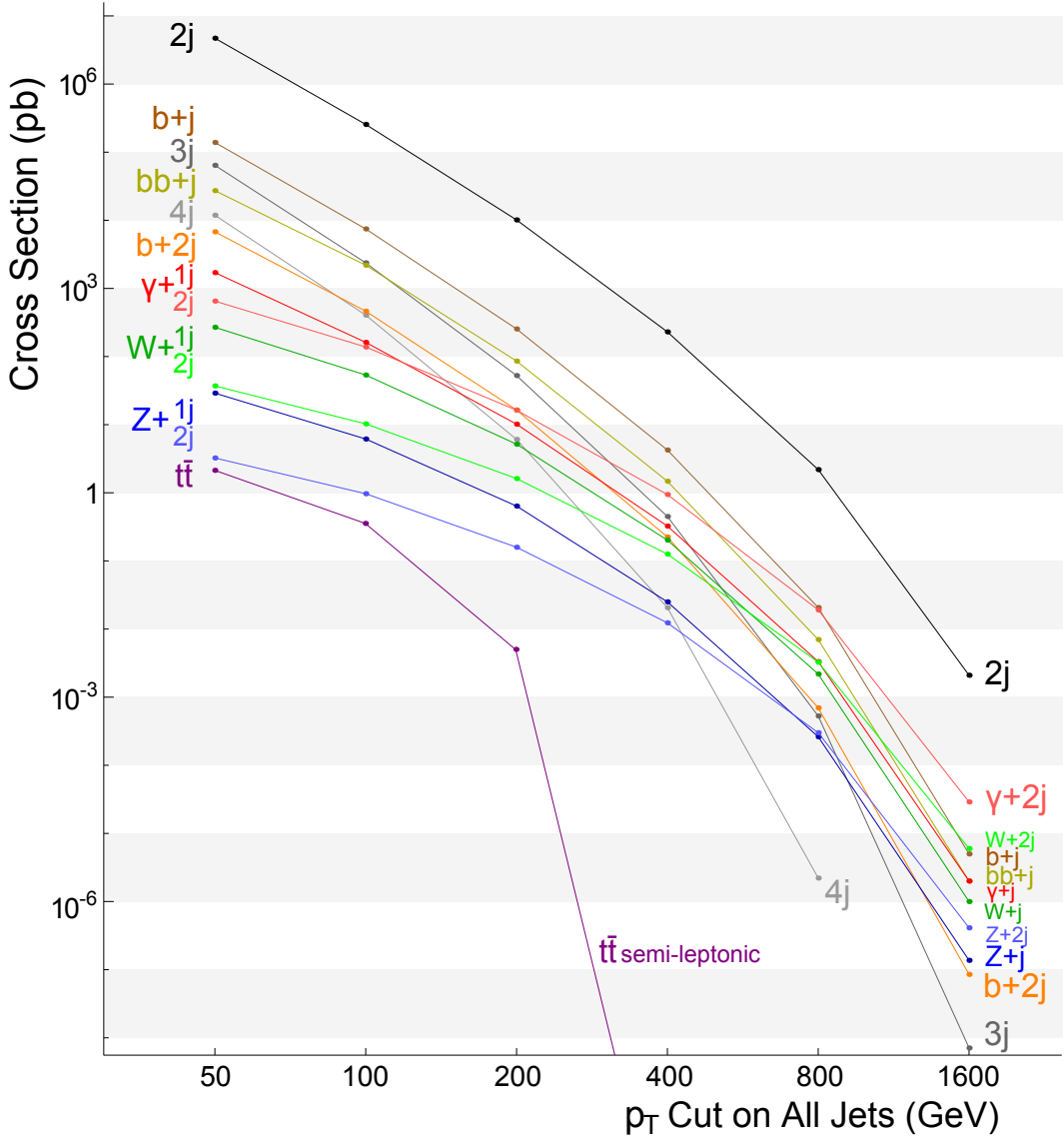}
\caption{
Leading order cross sections, including kinematic cuts and
branching ratios for $Z/W$ decay to include an electron or muon.
The $x$-axis indicates the $p_T$ cut applied to
\emph{all} light quarks and gluons, but not $b$-quarks.
The constraint on the $p_T$  for
$b$'s, photons, and charged leptons or neutrinos from $Z/W$ (though
not the $Z/W$ itself)
is fixed at 20\,GeV.
Note that the 3-jet cross section falls below $b$+2jets due to the harder
cuts on the non-$b$ jets.
The $t \bar t$ cross section refers to the semi-leptonic sample,
and, in contrast to all the other samples,
the $p_T$ cut is applied to only one of the two light-quark jets. Since its cross section
is so low, it will not be considered further.
}
\label{fig:Cross_Section_in_pb}
\end{center}
\end{figure}

The starting cross sections are shown in
Figure~\ref{fig:Cross_Section_in_pb}, as a function of
the $p_T$ cut applied to all light
quarks and gluons. along with the other cuts listed above.
If a sample has a bigger starting cross
section, it will be able to suffer harder purification cuts
while retaining a substantial number of events. In this plot, the
$t\bar{t}$ sample includes the semi-leptonic branching ratios
(2 leptons, 2 $b$'s and 2 light quarks) and has the $p_T$ cut
applied to only one of the light quark jets. Despite this
looser cut, the cross section drops precipitously above
200\,GeV, mostly due to the requirement that the jets be
separated by $\Delta R \ge 1$. Since the semi-leptonic
$t \bar{t}$ cross section is very small compared to the other
processes, we conclude that $t\bar{t}$ events are not a good
way to get a large quark jet sample, despite the fact that jets
coming from the hadronic $W$ decay are 100\% quark.

Instead of putting a cut on the $p_T$ of all the jets, we also
tried sorting jets by their rapidity. For example, we asked how
often the most (or least) central jet is initiated by a
particular parton. This was never more effective at
purification than sorting by $p_T$. Rapidity differences will
be used to further purify the samples, but for the starting
distributions, we stick with the $p_T$ cut.

\begin{figure}
\begin{center}
\includegraphics[width=0.32\textwidth]{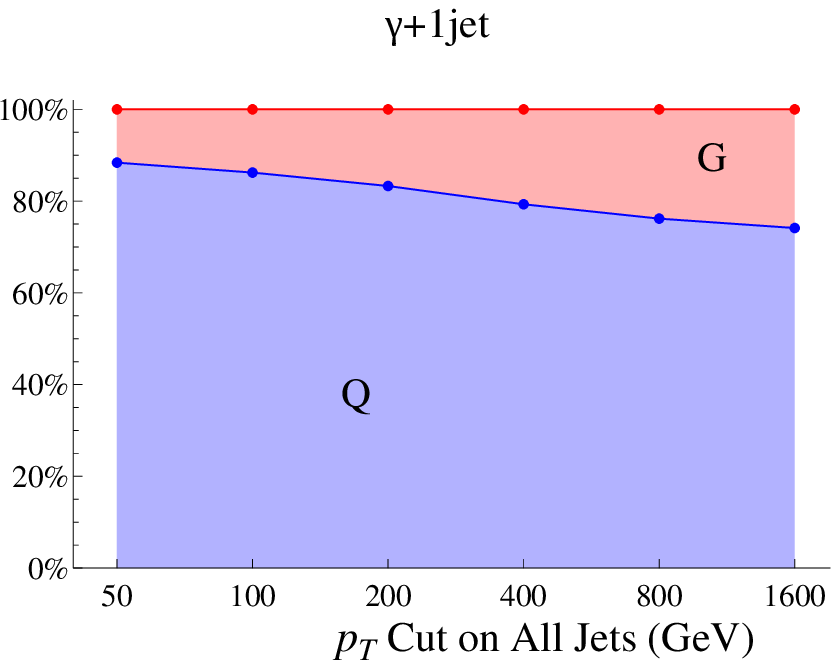}
\includegraphics[width=0.32\textwidth]{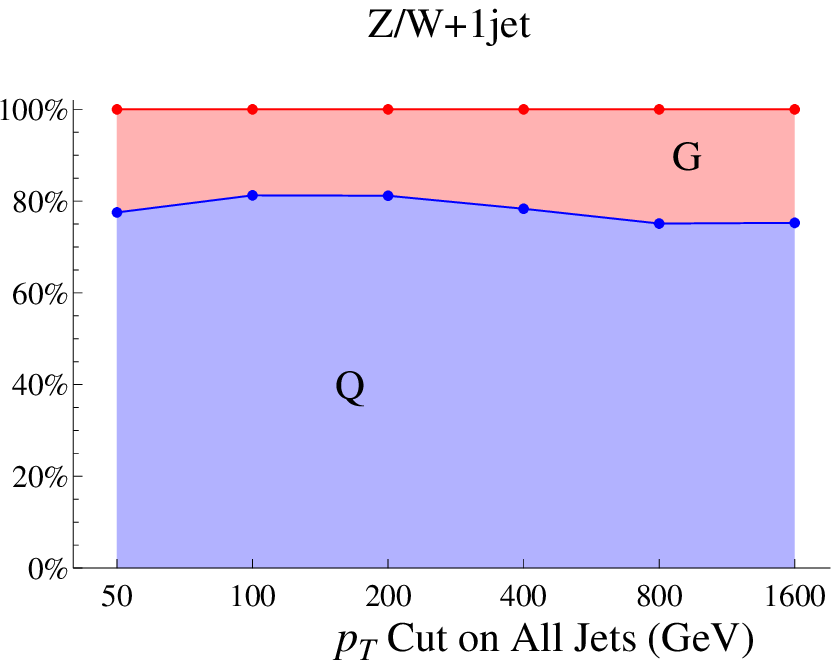}
\includegraphics[width=0.32\textwidth]{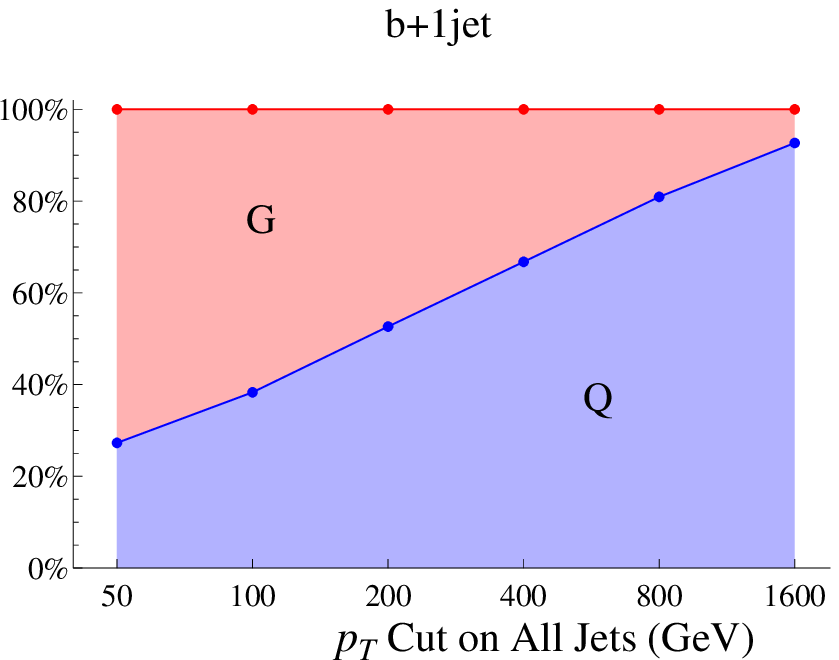}
\caption{
Fraction of $X$+1jet events where the jet is $uds$ quark (bottom and blue
in each plot) as compared to gluon (top and red).  The horizontal axis
is a $p_T$ cut on the jet, which in these events translates into an identical
$p_T$ cut on the other object.
}
\label{fig:1jet_fractions}
\end{center}
\end{figure}
\begin{figure}
\begin{center}
\includegraphics[width=0.32\textwidth]{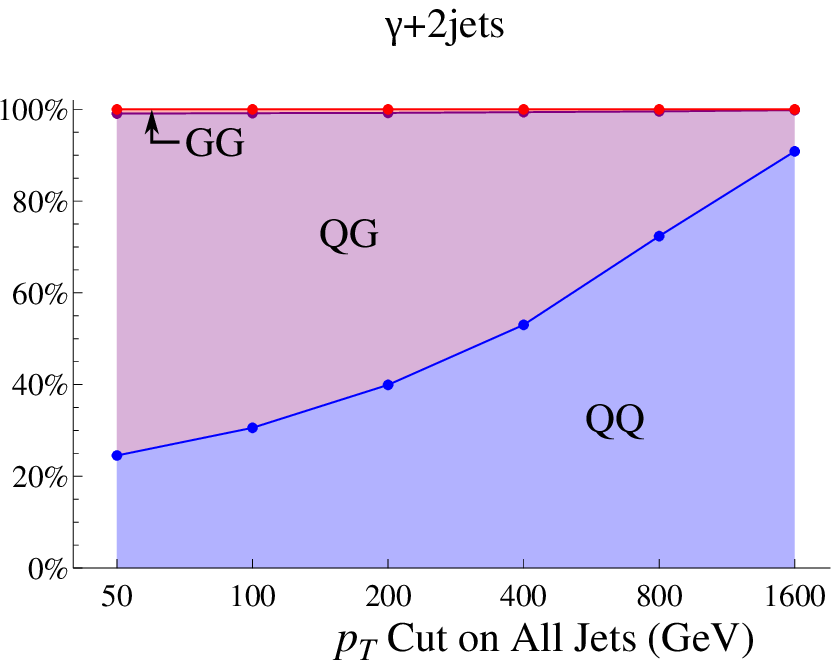}
\includegraphics[width=0.32\textwidth]{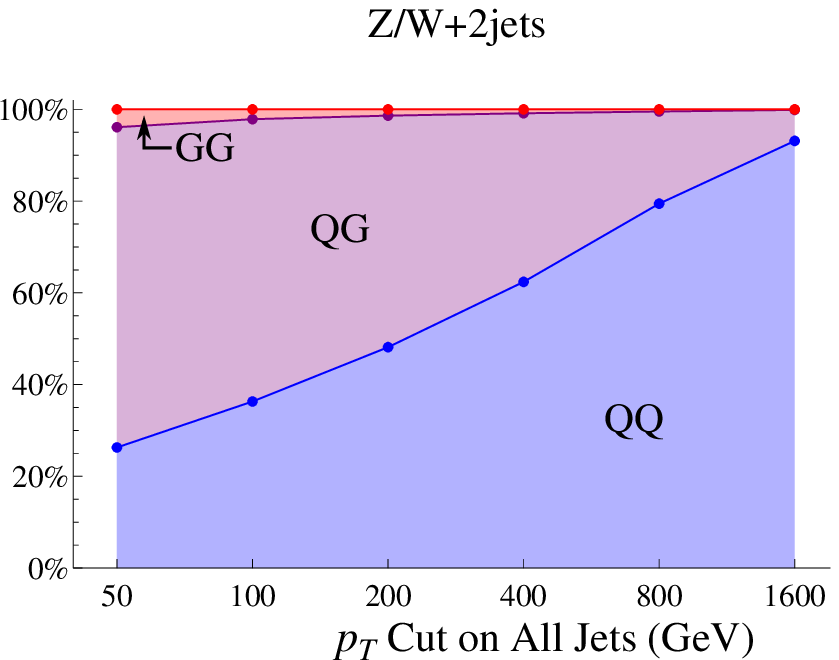}
\includegraphics[width=0.32\textwidth]{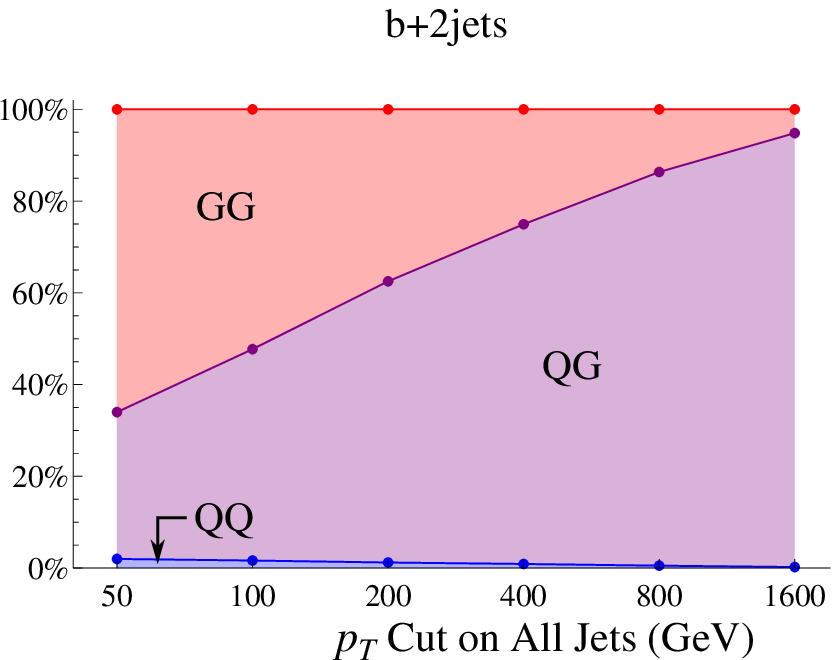}
\caption{
Fraction of $X$+2jet events where the jets are both light quark `QQ' (bottom blue)
vs one light quark one gluon `QG' (middle purple)
vs both gluon `GG' (top red).
Notice $\gamma+GG$ almost never happens, nor does $b+QQ$.
These are starting points for quark and gluon purification.
The horizontal axis is a $p_T$ cut on all jets, while the other
objects ($b$, $\gamma$, and leptons from $Z/W$)
have $p_T>20$\,GeV.
}
\label{fig:2jet_fractions}
\end{center}
\end{figure}
\begin{figure}
\begin{center}
\includegraphics[width=0.32\textwidth]{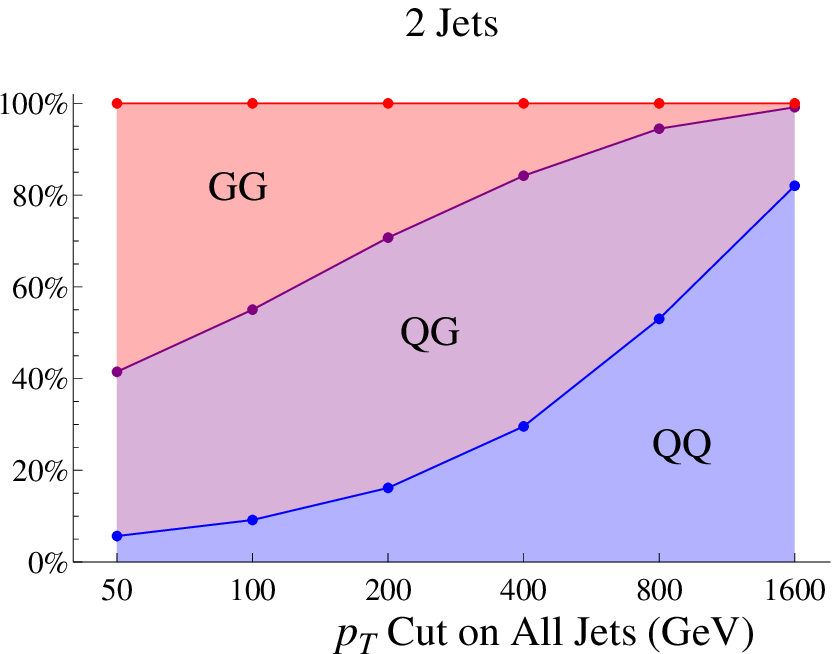}
\includegraphics[width=0.32\textwidth]{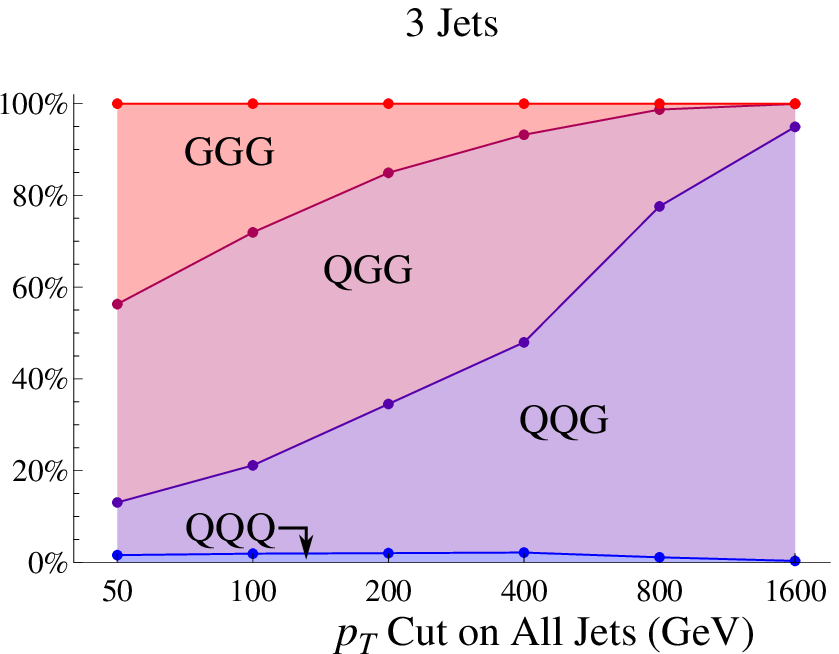}
\includegraphics[width=0.32\textwidth]{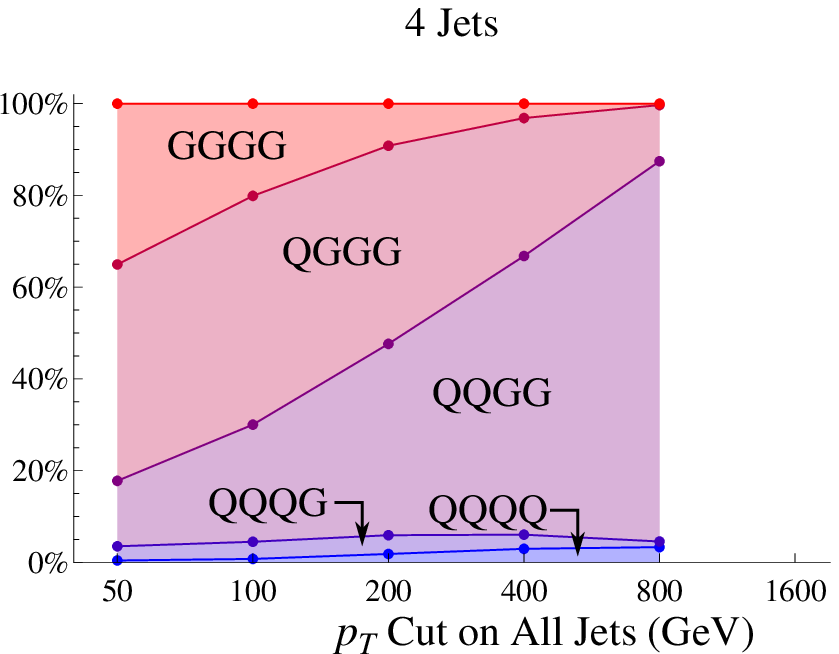}
\caption{
Division of the multijet (dominantly QCD) sample.
The horizontal axis is a $p_T$ cut on all jets.
Notice that all three jets are almost never all quark, and
in the 4-jet sample, there are almost always at least two gluons.
The 3-jet sample will be a staring point for gluon purification.
\label{fig:multijet_fractions}
}
\end{center}
\end{figure}

In the following, `quark jet' will always mean only $u$, $d$,
and $s$ quarks. Any $b$'s and $c$'s are treated as perfectly
taggable, although it is straightforward to put in the tagging
efficiencies. In Figures~\ref{fig:1jet_fractions} through
\ref{fig:multijet_fractions}, we show the fraction of quarks
and gluons produced in the various samples as a function of
$p_T$. When dijet events are referred to as `QG', that means
one jet is a gluon and the other is always a $uds$ quark. The
fraction of dijet events that are `QG' does not include cases with $b$ or
$c$ jets in the numerator or the denominator.

\begin{figure}
\begin{center}
\includegraphics[width=0.45\textwidth]{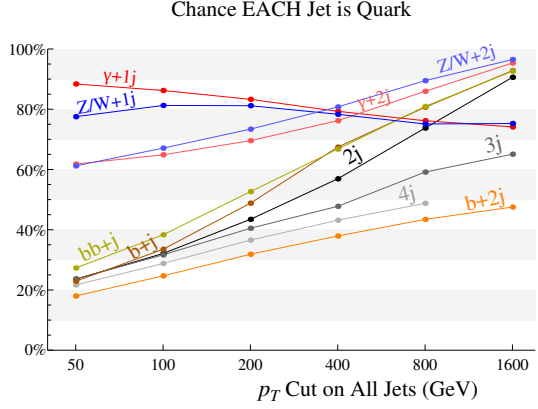}
\vspace{-0.1in}
\caption{
The chance that a given jet is a
light quark jet rather than a gluon jet.
(This ratio does not include bottom or charm.)
The $W$ and $Z$ were nearly identical and combined on this plot,
but they are slightly different from the photon, mostly due to the $\gamma$
and lepton cuts.
\label{fig:Chance_EACH_Jet_is_Quark}
}
\end{center}
\end{figure}

In Figure~\ref{fig:Chance_EACH_Jet_is_Quark}, we show the
probability that a given jet is a quark or gluon as a function
of $p_T$ for the different samples, assuming one jet is picked
at random. We see that $\gamma$+1jet or $W/Z$+2jets are good
for quark jets, and $b$+2jets or the 3 or 4-jet samples are
good for gluon jets. Again, this is just for the generic cuts
listed above, and we have not yet attempted to purify the
samples using rapidity or other kinematic information.

\begin{figure}
\begin{center}
\includegraphics[width=0.45\textwidth]{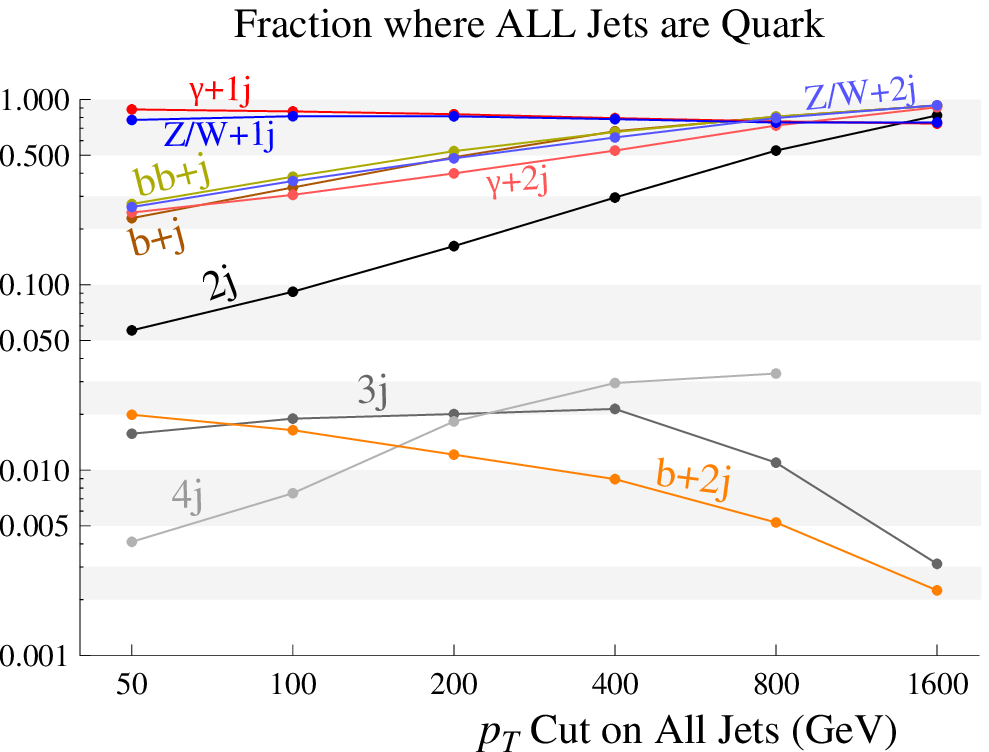}
\qquad
\includegraphics[width=0.45\textwidth]{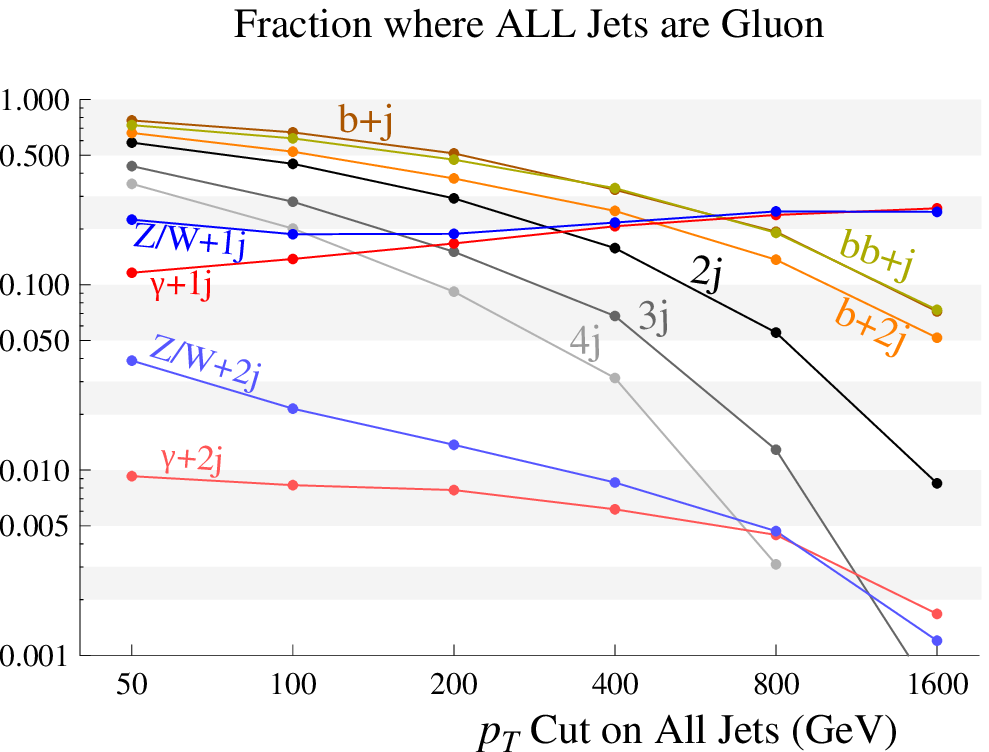} \\
\qquad \\
\includegraphics[width=0.45\textwidth]{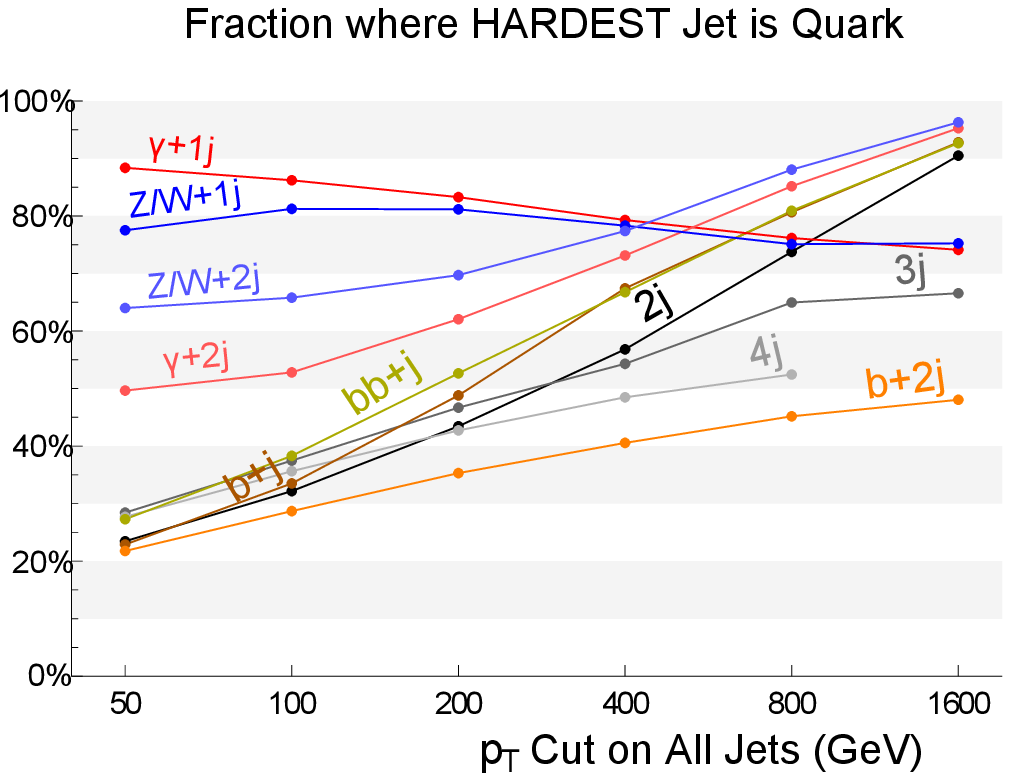}
\qquad
\includegraphics[width=0.45\textwidth]{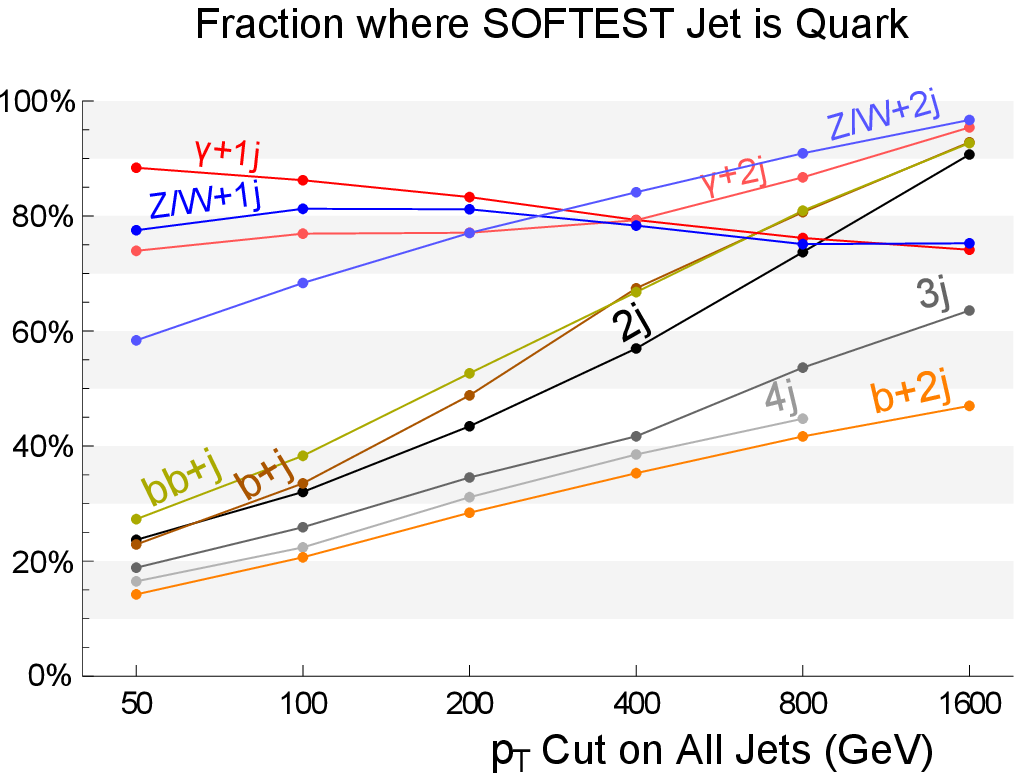}
\vspace{-0.1in}
\caption{
The top row shows the fraction of events  where \emph{all} jets are quark or gluon, on a log scale.
The bottom row shows the fraction where the \emph{highest} $p_T$ jet is quark,
and where the \emph{lowest} $p_T$ jet is quark, on a linear scale.
(One minus this fraction are gluon jets.)
Having more jets allows for more kinematic handles and potentially better purity.
\label{fig:Fraction_where_HIGHEST}
}
\end{center}
\end{figure}

In order to purify the samples, we can go two ways. One
approach is to reject events so that all of the jets in the
remaining events have either all quark jets or all gluon jets.
In the top panels of Figure~\ref{fig:Fraction_where_HIGHEST},
we show the fraction of events where \emph{all} jets are quark or gluon. Note
that the vertical axis in these plots is logarithmic.
The other
approach is to look at particular jets in an event, eventually
hoping to apply kinematic cuts to purify the quark or gluon content of
\emph{that} jet. (Such cuts are the topic of the next section.) In the bottom of
Figure~\ref{fig:Fraction_where_HIGHEST}, we show the fraction
where the hardest or softest jet is quark.
These starting points indicate that quark jets will be
easier to purify than gluon jets.

\clearpage

\pagebreak

\clearpage




\clearpage

\pagebreak

\section{Purifying the samples \label{sec:pure}}
In this section, we consider how to improve the purity by
judicious kinematic cuts. It's actually quite challenging to get high purities, as we will see.
For example, if you start with a 50\% pure quark
sample and you find a set of cuts that reject two gluons for
every quark kept, your new purity is \emph{not} 75\%, but only
66\%.  To reach 75\%, you need a cut that rejects three gluons
for every quark.

Any cut will have some efficiency  $\eps_q$ to
keep quark jets and a different efficiency $\eps_g$ to keep
gluon jets.
Let $q$ be the starting fraction of events
where the jet in question (e.g. the lower $p_T$ `softer' jet)
is a light quark, and $g=1-q$ the fraction of events where it
is a gluon. Then, after a cut,
\begin{equation}
q
\ =
\frac{q}{q+g}
\ \ra{cut} \
\frac{q \eps_q}{q \eps_q + g \eps_g}
\ = \
1 \left/ \left(1+\frac{g}{q}\frac{\eps_g}{\eps_q} \right) \right.
\ = \
q_\mathrm{new}
\end{equation}
Say we want to optimize the quark purity. One particular cut on
the set of kinematic variables will be the best cut for a
\emph{particular} quark efficiency $\eps_q$. This will be the
cut that lowers the gluon acceptance $\eps_g$ as much as
possible.


To reach a given quark purity, it obviously helps to start with
a sample that's mostly quark.  But it is possible to find
effective kinematic cuts that improve a mediocre quark
purity.  This is the case in the $\gamma$+2jet sample.
Strong cuts can increase the quark purity
quite a bit for some samples, but at the cost of a much
lower cross section. In the following, we will be careful to
express our results as the cross section for quark and gluon
jets with a given purity.

\subsection{Quark jet purification}
We begin by discussing purifying quark jets. As can be seen
in Figure~\ref{fig:1jet_fractions}, the $\gamma$+1jet
sample appears to be a good starting point, with roughly 80\%
quarks. This fraction is just the fraction of direct photons produced in the
annihilation channel $q\bar{q} \to g \gamma$ (20\%) versus the
Compton channel $q g \to q \gamma$ (80\%), which is in turn set
by the gluon and $\bar{q}$ PDFs. Since the gluon PDF is larger
than the $\bar{q}$ PDF in a proton, the Compton channel
dominates. Unfortunately, the 1-jet samples, such as
$\gamma$+1jet or $W/Z$+1jet, do not leave many options for kinematic
cuts. Rapidity cuts do not do much, since at high $p_T$, the
jets are more-or-less central, and the cross sections are
basically fixed by the PDFs. In fact, the quark purity saturates at roughly 88\%.
Thus, is helpful to have
additional jets to get an additional handle on the kinematics, which will lead us
to purities approaching 100\%.

\begin{figure}
\begin{center}
\vspace{-0.1in}
\psfrag{j0eta}{\quad $\eta_{j_1}$}
\psfrag{j1eta}{\quad $\eta_{j_2}$}
\psfrag{a0eta}{\quad $\eta_\gamma$}
\includegraphics[width=0.3\textwidth]{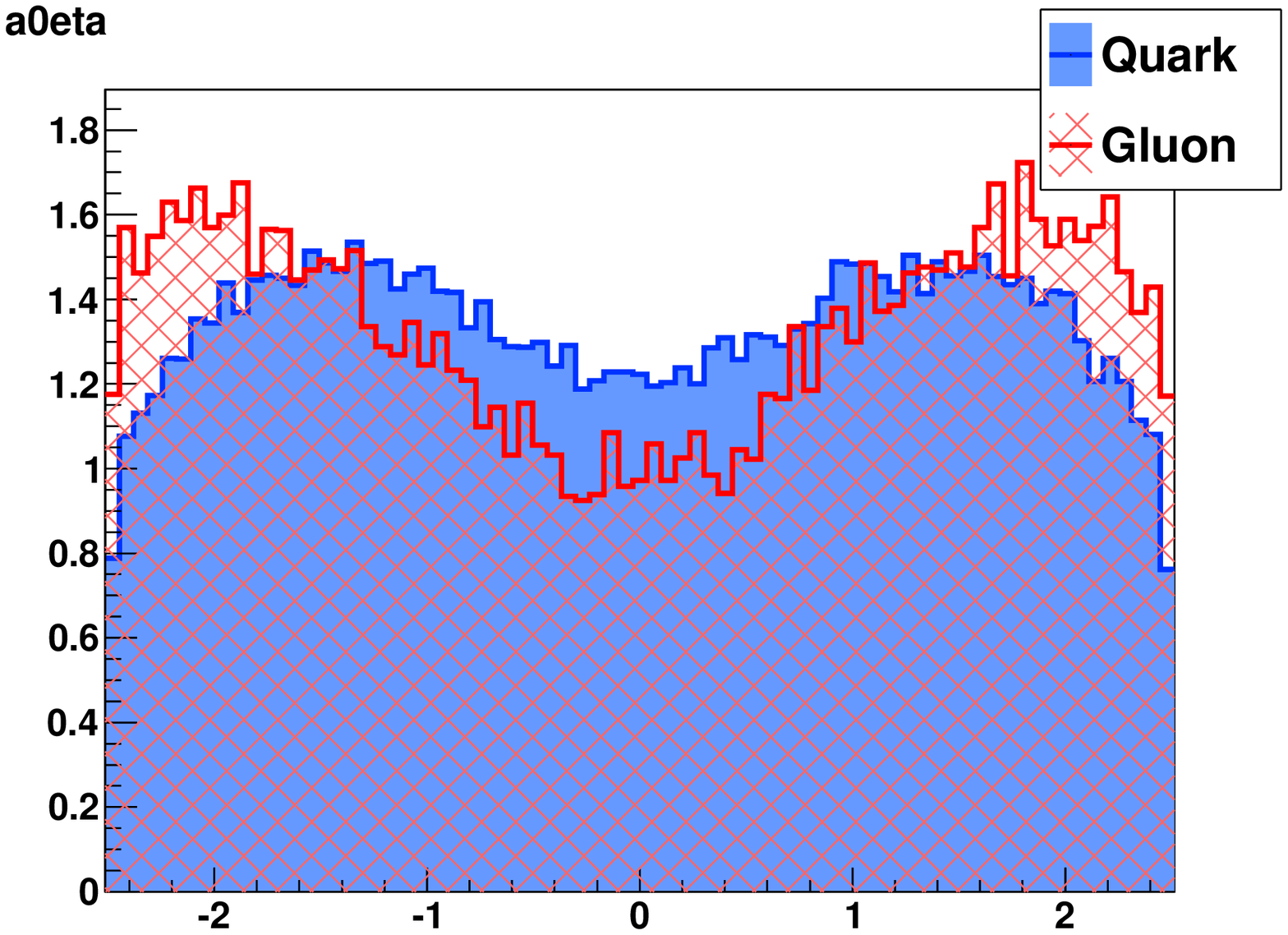}
\includegraphics[width=0.3\textwidth]{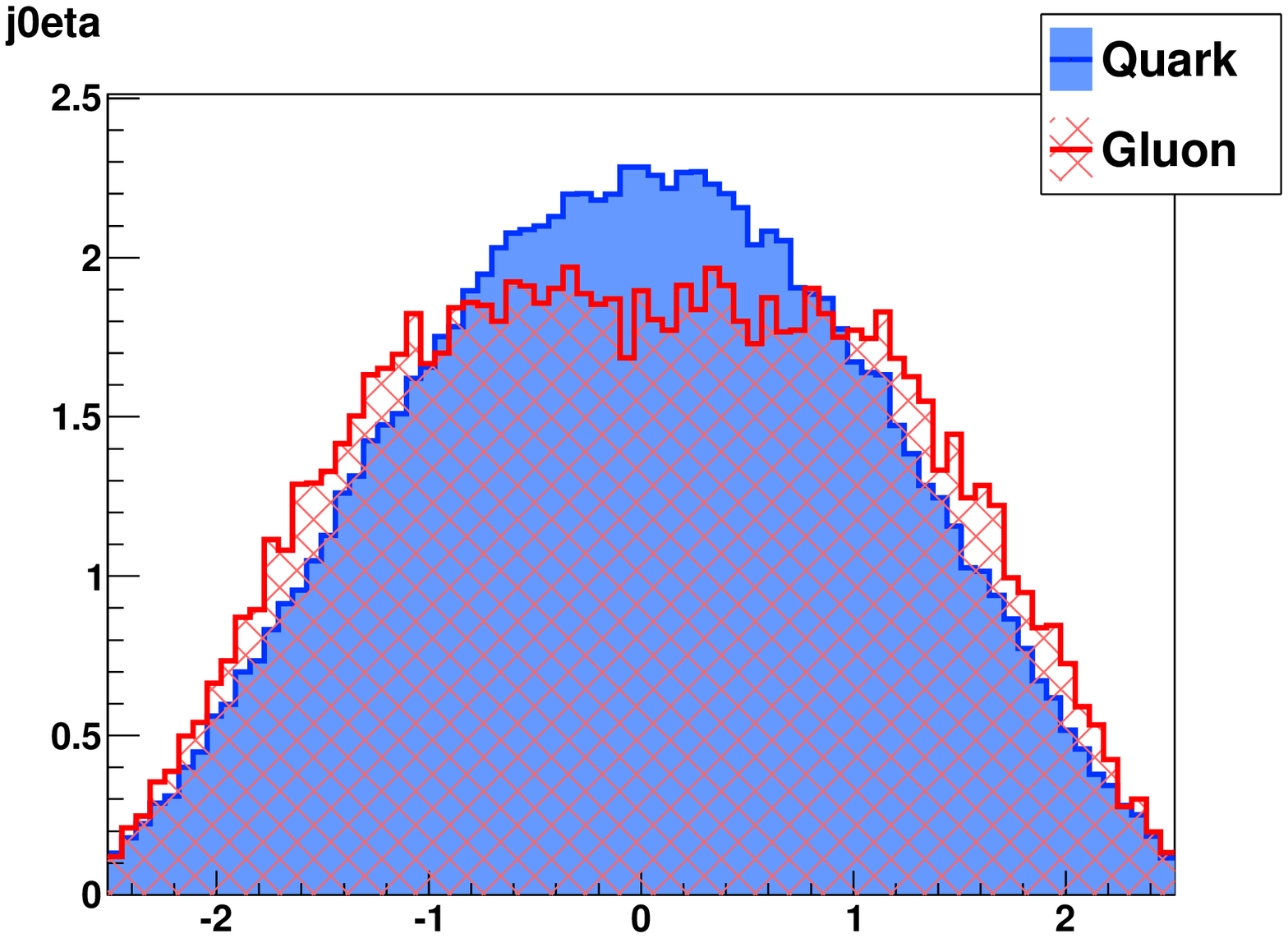}
\includegraphics[width=0.3\textwidth]{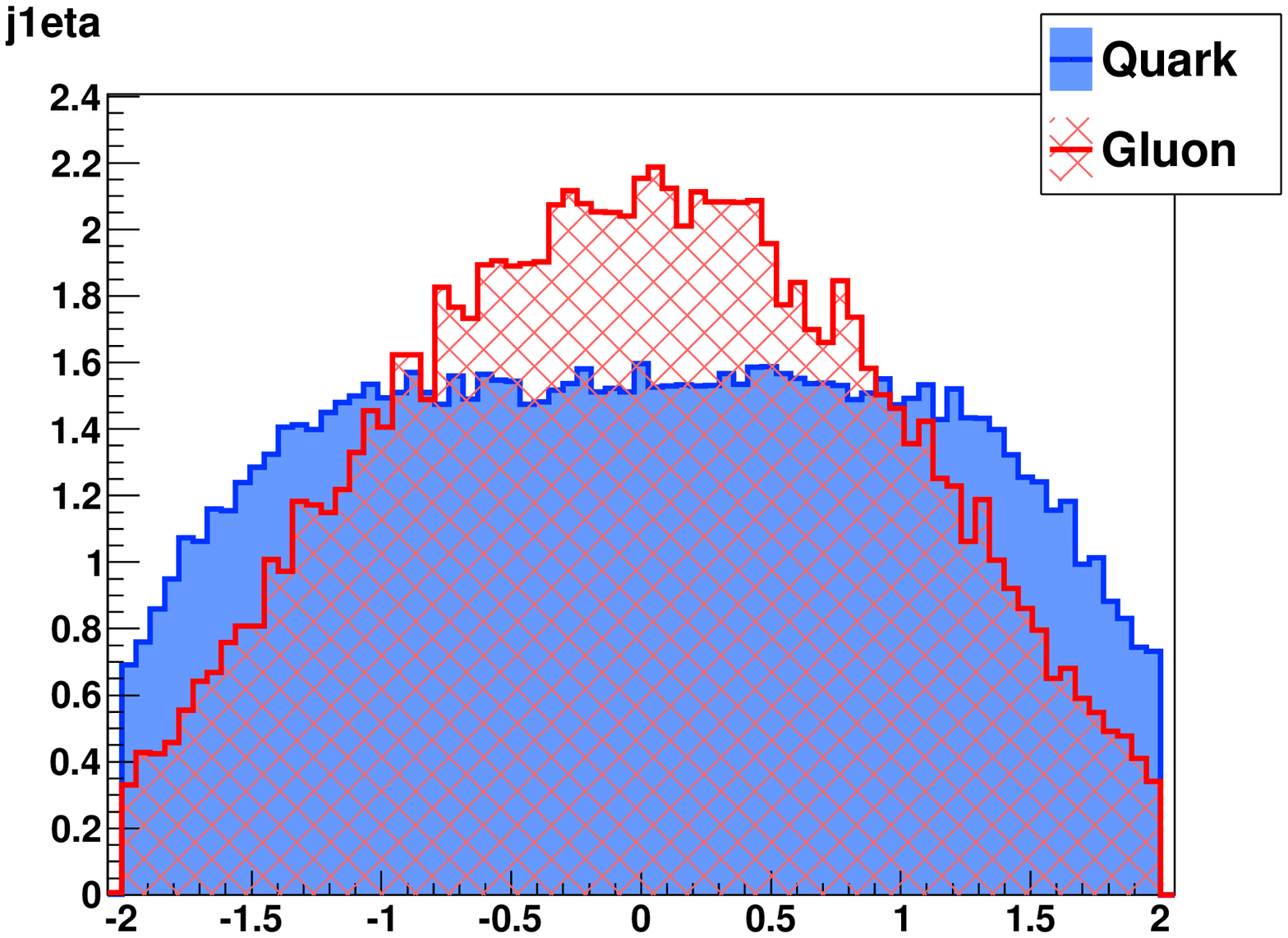}
\vspace{-0.1in}
\caption{
To purify quarks, the best starting point is the softer jet in the $\gamma$+2jet sample.
The $\eta$ of the photon (\textbf{left}) along with the harder (\textbf{center})
and softer (\textbf{right}) jets look different when the
\emph{softer} jet is a quark (blue solid) vs a gluon (red hashed).
These distributions are normalized to equal area.
(200\,GeV sample shown)
}
\label{fig:quark1d}
\end{center}
\end{figure}

\begin{figure}
\begin{center}
\psfrag{Signal}{\qquad  \quad \quad Quark} 
\psfrag{Background}{\qquad  \quad \quad Gluon}  
\psfrag{a0eta_j0eta}{\qquad \quad  Likelihood}
\psfrag{j0eta}{\small $\eta_{j_1}$}
\psfrag{j1eta}{\small $\eta_{j_2}$}
\psfrag{a0eta}{\small $\eta_\gamma$}
\includegraphics[width=0.3\textwidth]{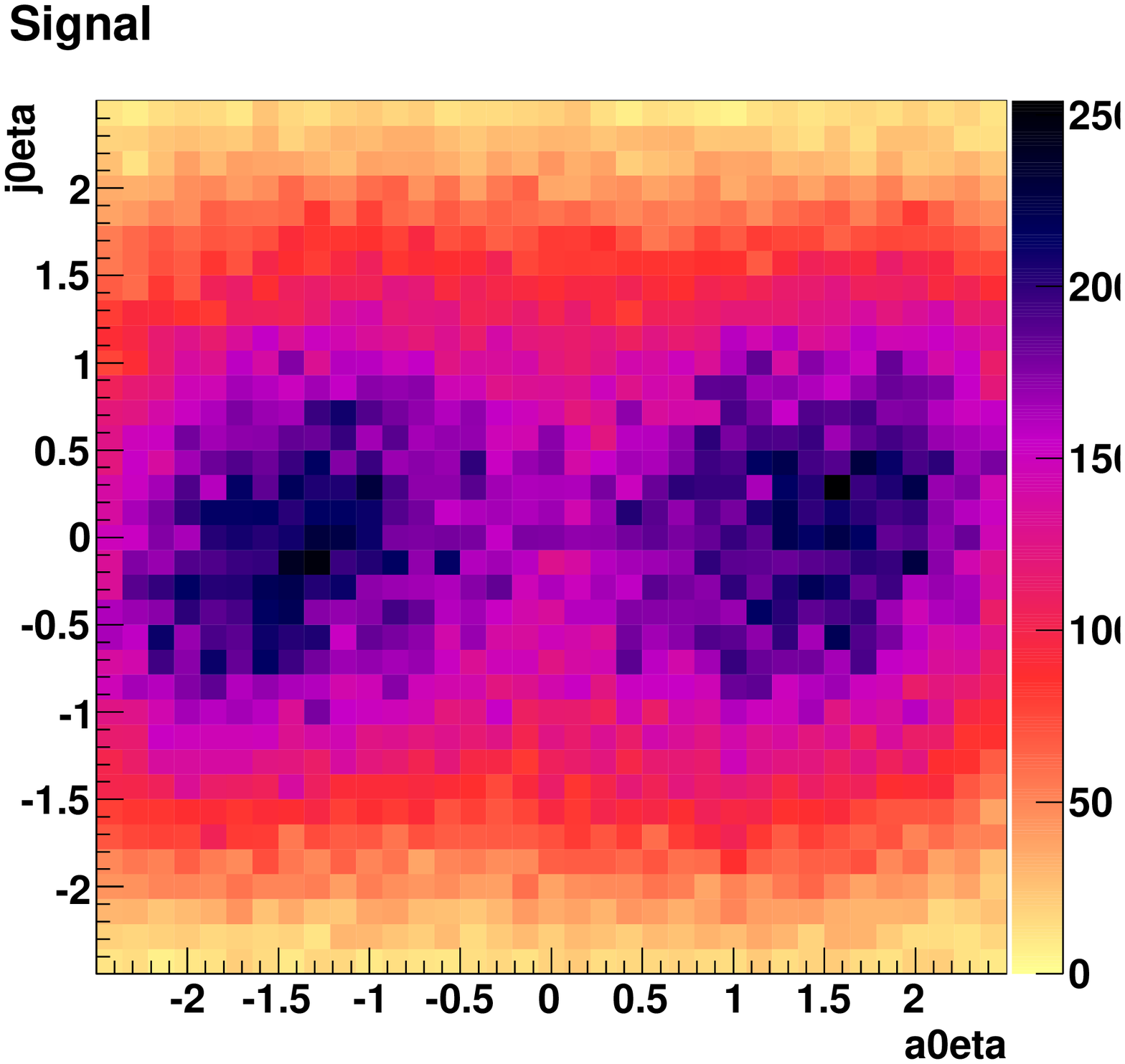}
\includegraphics[width=0.3\textwidth]{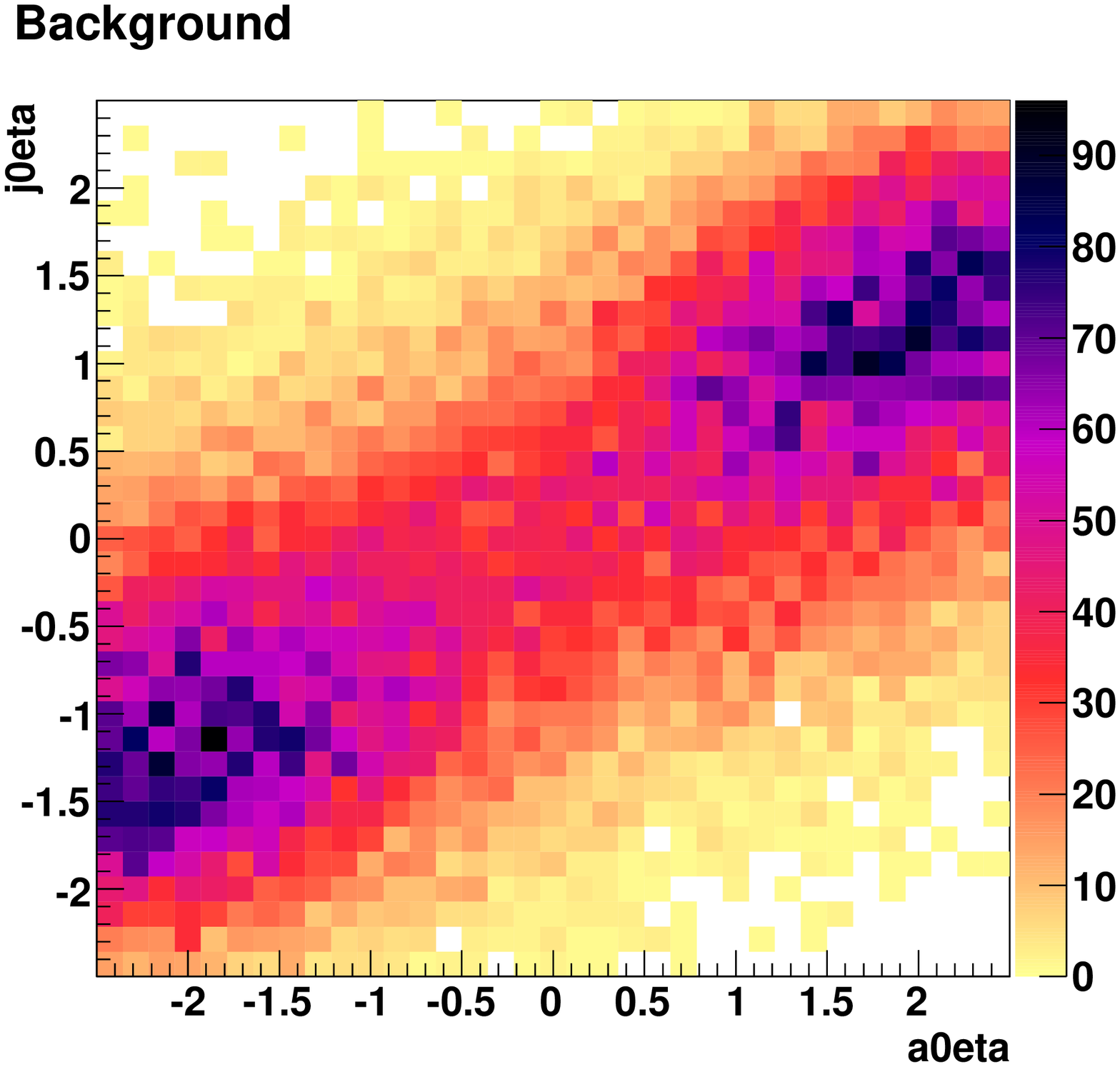}
\includegraphics[width=0.3\textwidth]{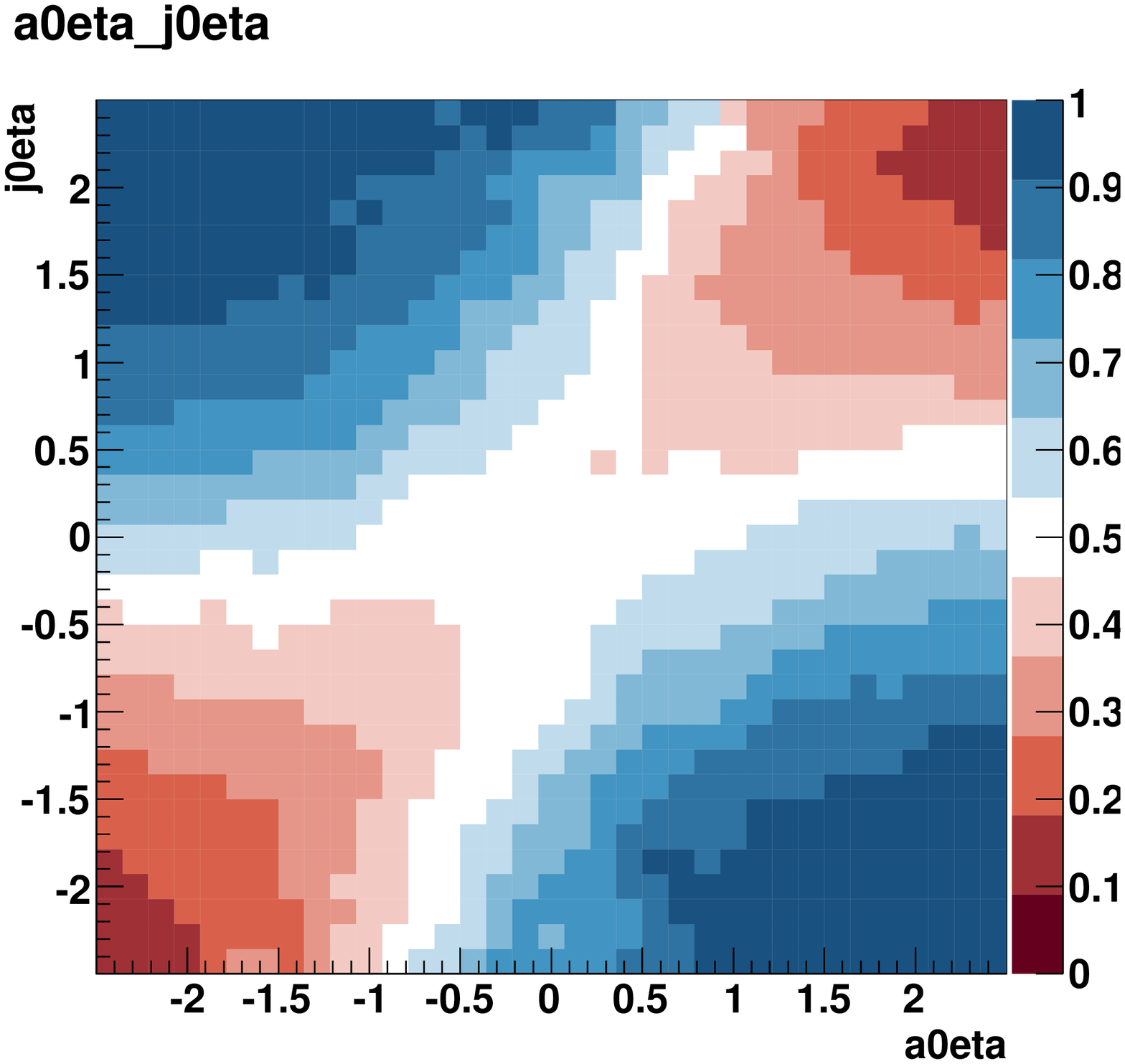}
\vspace{-0.1in}
\caption{
For the quark-heavy $\gamma$+2jet sample,
a 2D version of last figure's first two histograms:
$\eta_\gamma$ of the photon vs $\eta_{j_1}$ of the harder jet.
The \textbf{left} histogram is for when the softer jet is a quark,
and the \textbf{center} histogram is for when the softer jet is a gluon.
Though we are trying to purify the \emph{softer} jet, it's best to
cut on $\eta_\gamma$ and $\eta_{j_1}$ of the \emph{harder} jet.
>From the left histogram it's clear that when the softer jet is a quark,
the harder jet is quite central and
the photon's $|\eta|$ is higher and uncorrelated.
When the softer jet is a gluon, the harder jet is often toward the edge
of our $\eta_j$ cut, with the photon nearby in $\eta$.
Correlations are lost of one takes the absolute value of these $\eta$s.
The likelihood ratio on the \textbf{right} combines each bin as $q/(q+g)$, with
blue being more quark-like.
When the photon and harder jet are widely separated in $\eta$, the
softer jet is likely quark.
(200\,GeV sample shown)
}
\label{fig:quark2d}
\end{center}
\end{figure}

\begin{figure}
\begin{center}
\psfrag{j0eta}{\quad  $\eta_{j_1}$}
\psfrag{j1eta}{\quad  $\eta_{j_2}$}
\psfrag{a0eta}{\quad  $\eta_\gamma$}
\psfrag{a0eta * j0eta}{\quad  $\eta_\gamma  \  \eta_{j_1}$}
\includegraphics[width=0.21\textwidth]{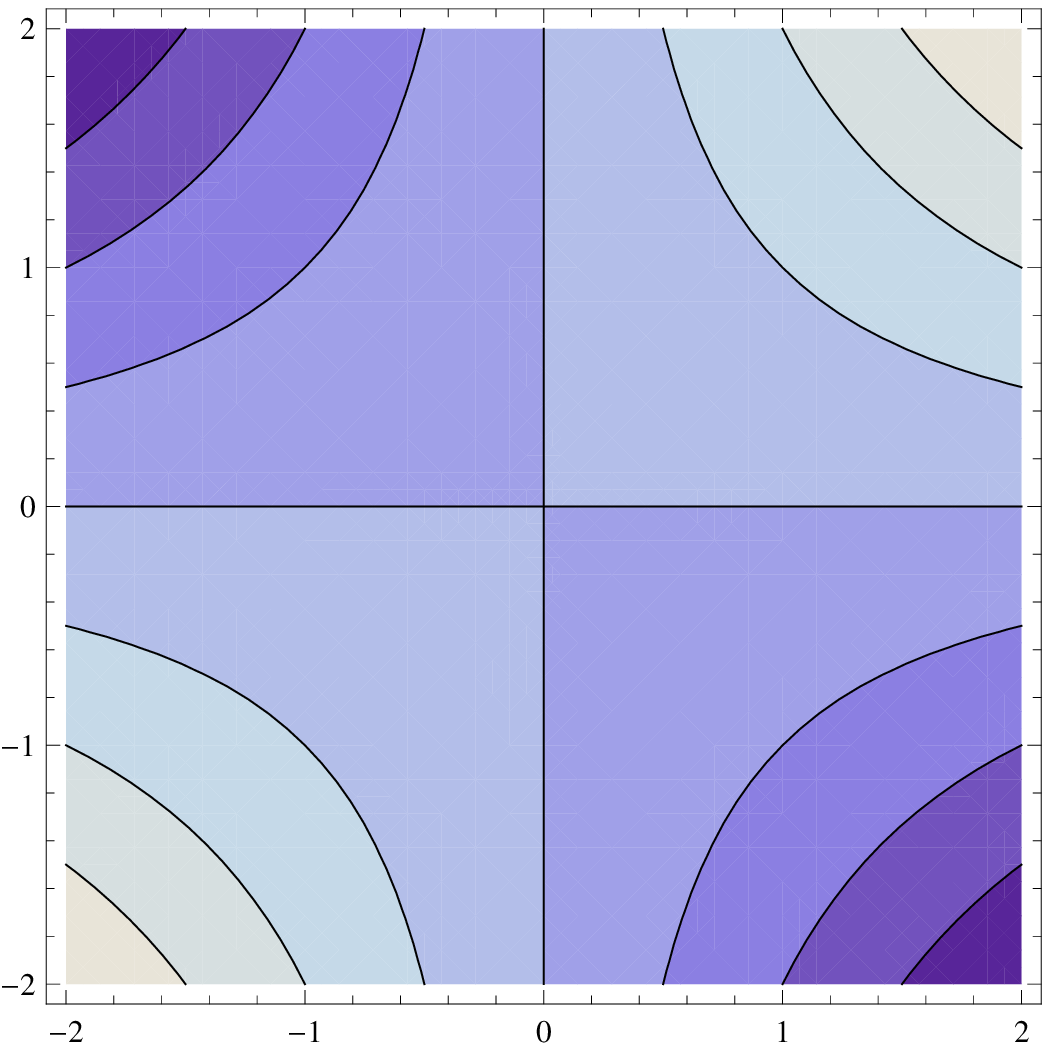}
\includegraphics[width=0.31\textwidth]{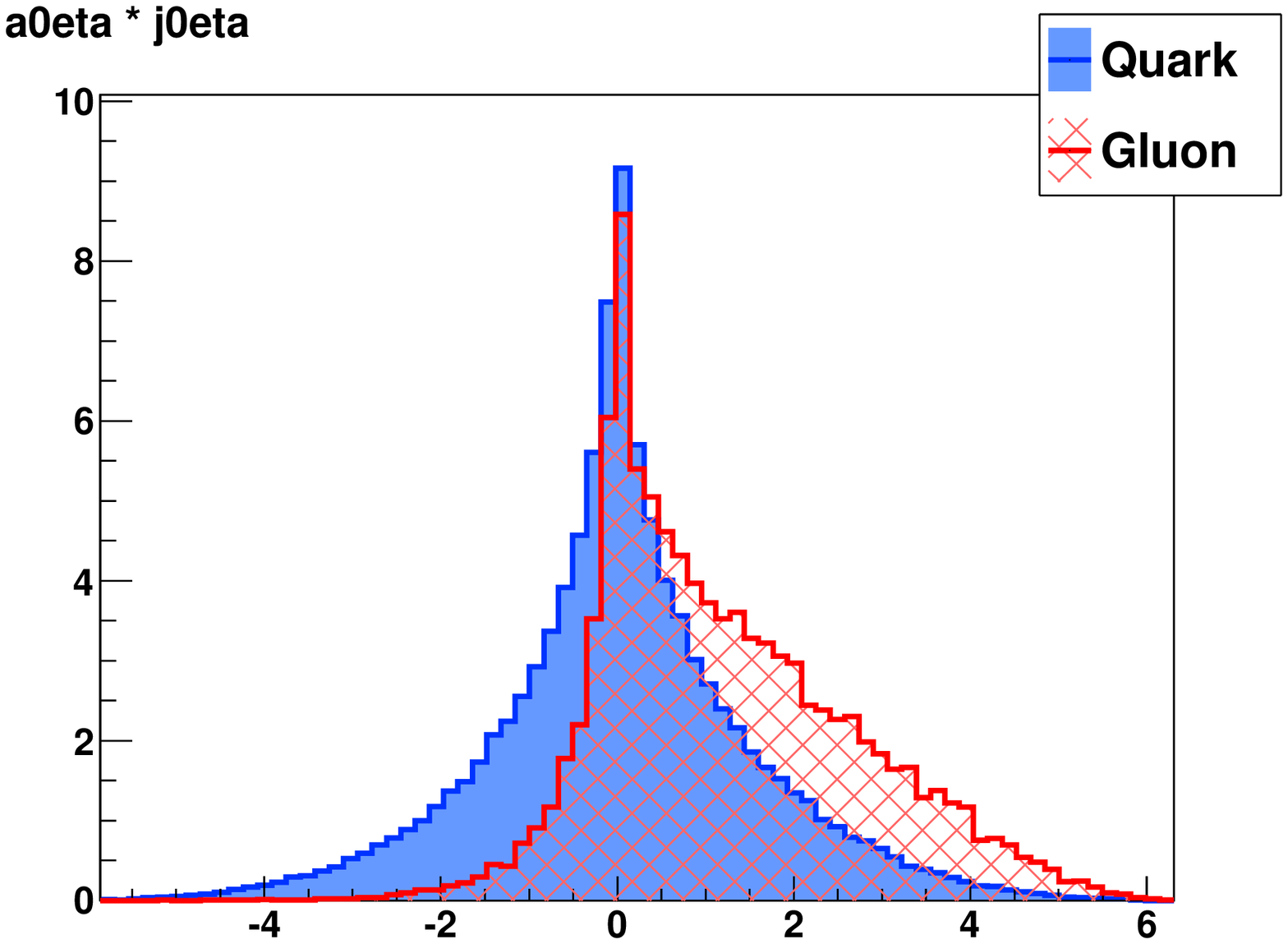}
\vspace{-0.1in}
\caption{
Cutting on any contour in
the 2D likelihood distribution above is statistically
the optimal discriminant for each quark efficiency,
given only these two variables.
The contours are roughly given by $\eta_\gamma \ \eta_{j_1}$,
which is plotted here on the \textbf{left}.
We will see that this single variable captures most of the discrimination
power of the full 9D multivariate likelihood estimate.
On the \textbf{right} is the distribution of this
product. (200\,GeV sample shown)
}
\label{fig:quark_a0eta_T_j0eta}
\end{center}
\end{figure}

We turn next to the the next best sample, $\gamma$+2jets. Note
that $W/Z$+2jets is kinematically very similar, but since it
has a smaller starting cross section, we focus on the photon.
The rapidity distributions for the photon and the softer and
harder jets in the samples are shown (for $p_T \gtrsim
200$\,GeV) in Figure~\ref{fig:quark1d}. These 1D distributions
look like they contain some information, but there is in fact
more information in their correlations.
Figure~\ref{fig:quark2d} shows the 2D distribution of the
rapidity of the harder jet and the rapidity of the photon. The
likelihood map constructed from these distributions is shown in
the third panel. Contours of constant likelihood are very well
approximated as contours where the product of the rapidities
$\eta_\gamma \eta_{j1}$ is constant, as shown in
Figure~\ref{fig:quark_a0eta_T_j0eta}. The quark/gluon
discriminant for this product variable is also shown in
Figure~\ref{fig:quark_a0eta_T_j0eta}. It clearly has more
discrimination power than any of the individual rapidities.

\begin{figure}
\begin{center}
\psfrag{dRaj0}{\quad  $\Delta R_{\gamma j_1}$}
\psfrag{dRaj1}{\quad  $\Delta R_{\gamma j_2}$}
\psfrag{a0j1mix}{\quad  $\eta_\gamma \  \eta_{j_1} + \Delta R_{\gamma j_2}$}
\includegraphics[width=0.32\textwidth]{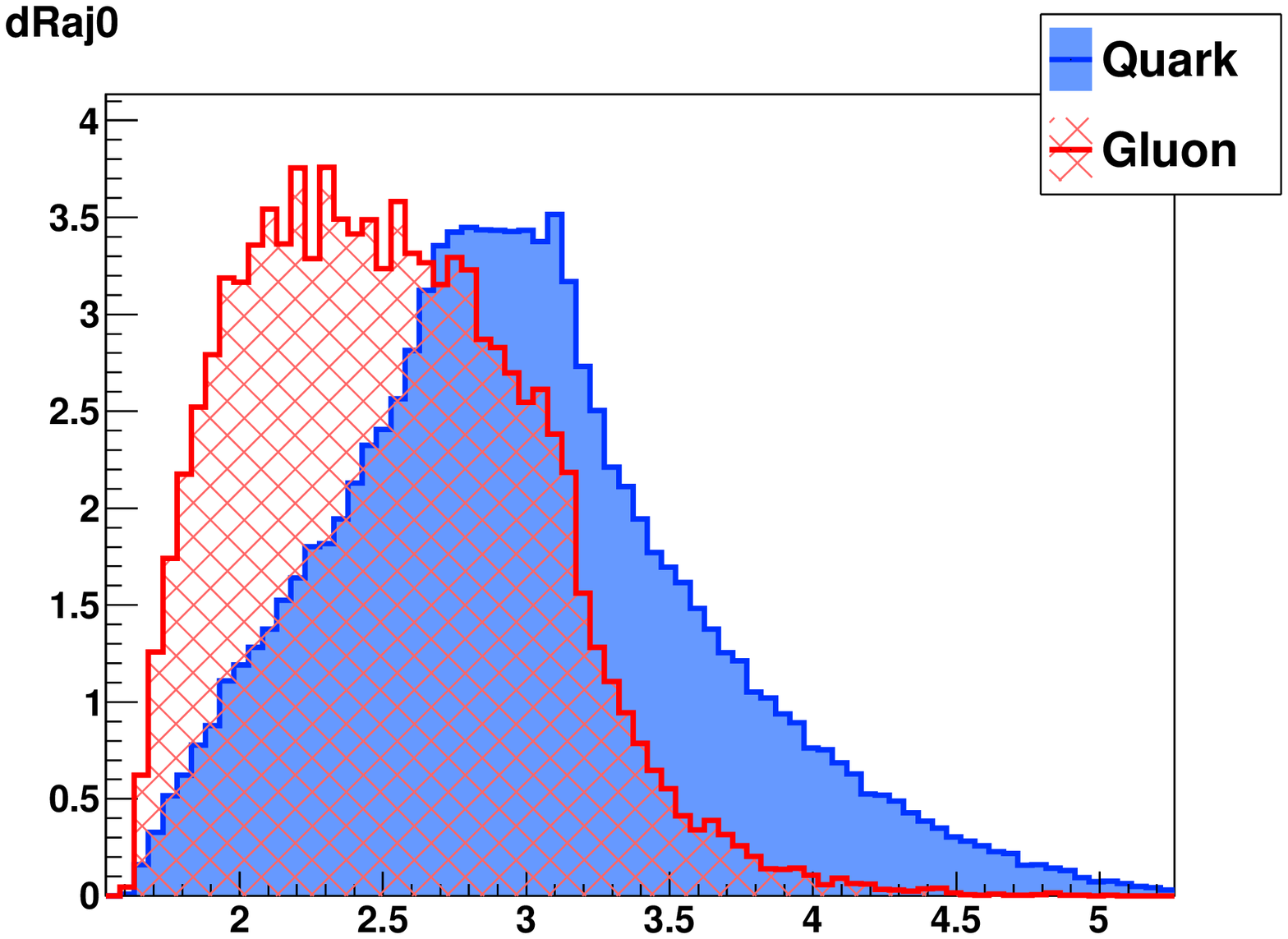}
\includegraphics[width=0.32\textwidth]{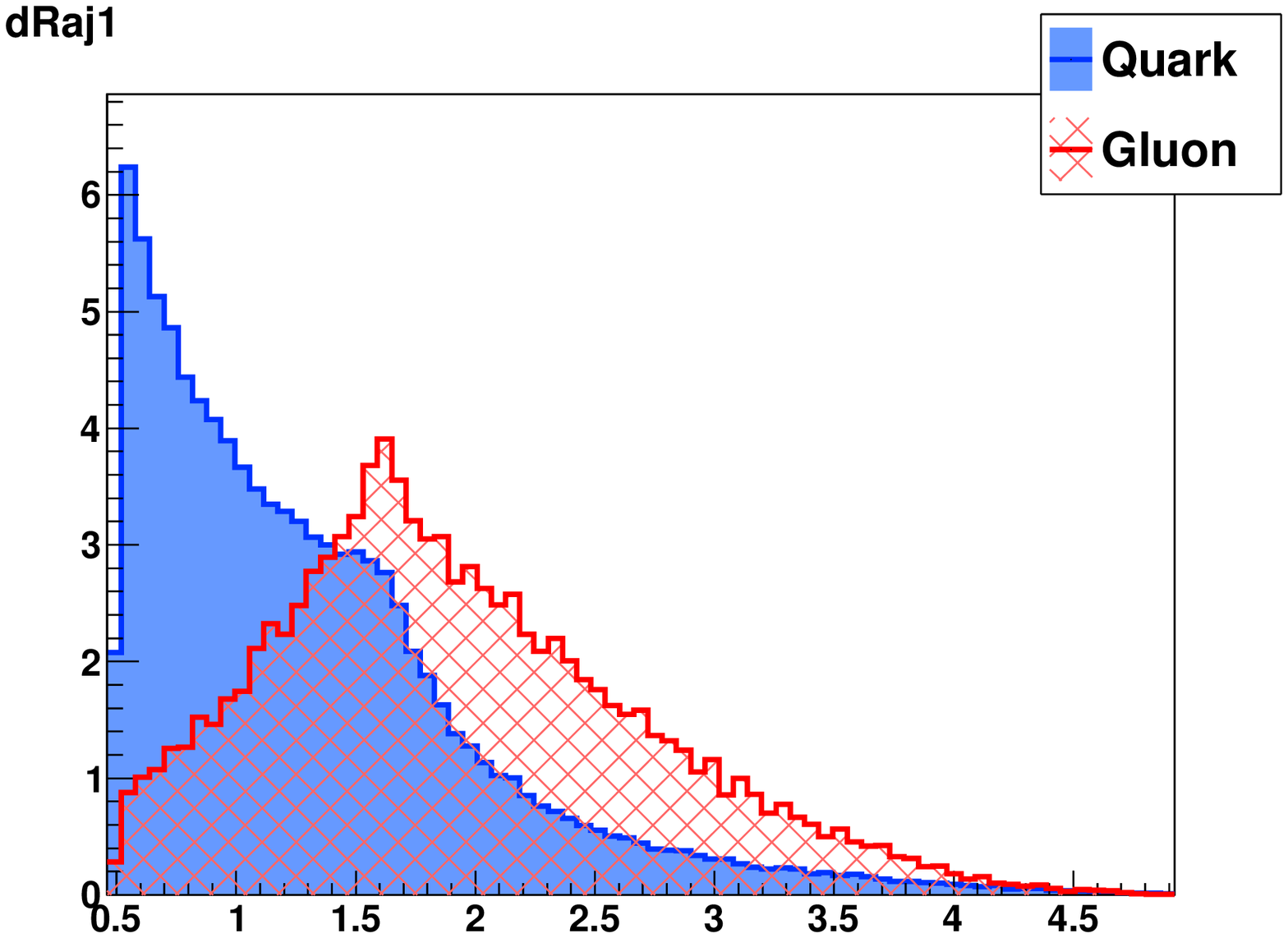}
\includegraphics[width=0.32\textwidth]{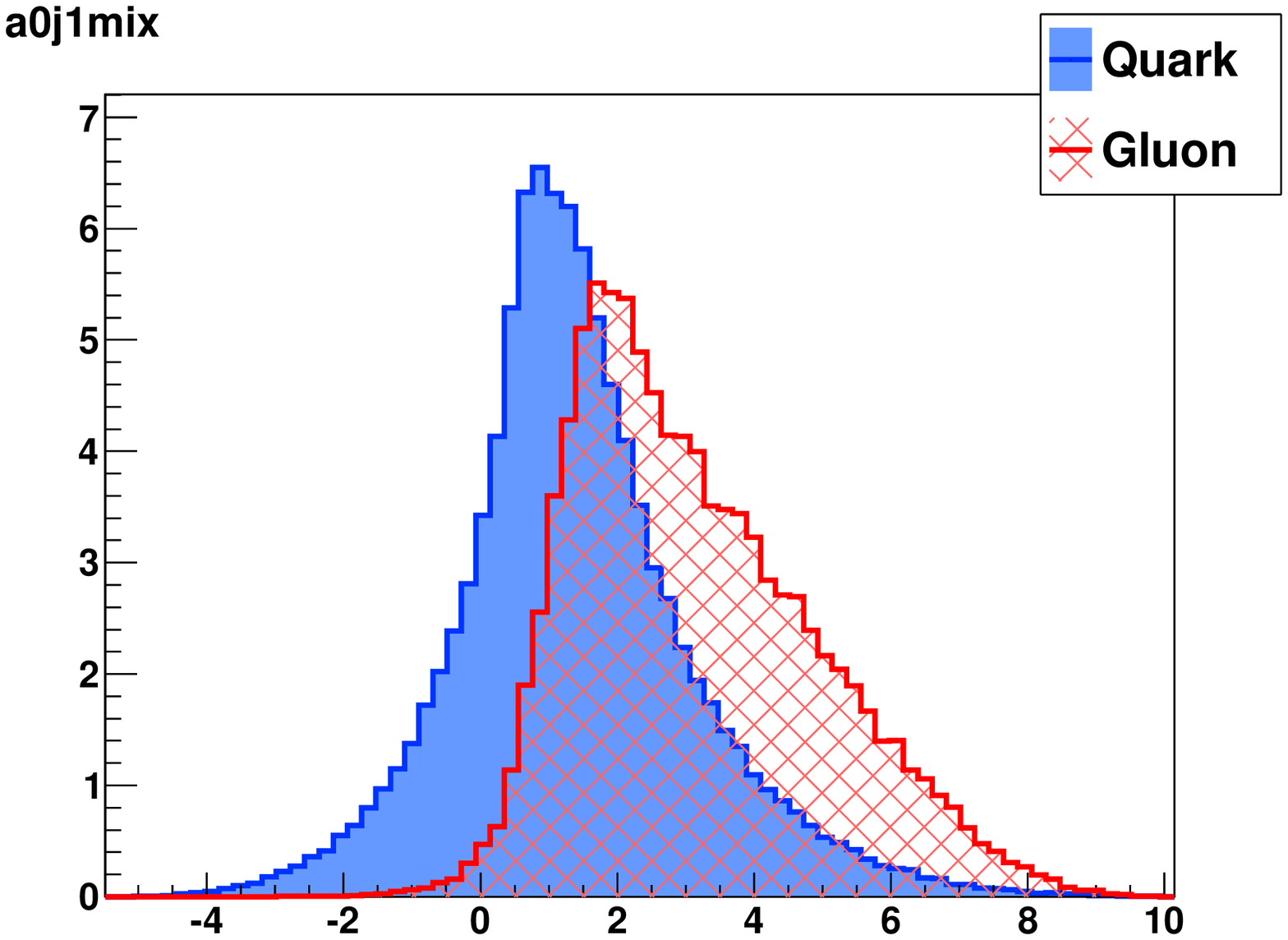}
\caption{
For the quark-heavy $\gamma$+2jet sample,
distance between the photon and the harder jet (\textbf{left})
and softer jet (\textbf{center}).  Notice that the photon is often
as collinear with the softer jet as our
$\Delta R_{\gamma j_2} > 0.5$ cut allows.
Doing the same 2D likelihood examination as before, an even
better single variable discriminant is:   $\eta_\gamma \  \eta_{j_1} + \Delta R_{\gamma j_2}$,
a combination of the product of $\eta$ of the photon and \emph{harder} jet,
plus the distance to the \emph{softer} jet.  The distribution of this
mixed variable is shown on the \textbf{right}. (200\,GeV sample shown)
}
\label{fig:quarkdR}
\end{center}
\end{figure}

Another option for the $\gamma$+2jet sample is to consider
the $\Delta R$'s between the photon and the jets. Due to a
collinear singularity in $q\to q \gamma$, it is natural to
expect the photon to be close to one of the quarks. This is in
contrast to the gluon case, since there is no $g \to g \gamma$ vertex.
The distribution of $\Delta R$ between the photon and
each jet is shown in Figure~\ref{fig:quarkdR}. Performing a
similar 2D likelihood analysis as with just the rapidity
inputs, we find the that the single variable
$\eta_\gamma \eta_{j 1} + \Delta R_{\gamma j_2}$ does very
well. Its distribution is also shown.

In constructing unusual variables like
$\eta_\gamma \eta_{j 1} + \Delta R_{\gamma j_2}$,
it is natural to wonder if
we are being sufficiently comprehensive. Considering that for a
sample with $n$ final-state on-shell quarks and gluons, there
are only $3n$ degrees of freedom,
it is possible simply to put
these 6, 9 or 12 variables into a multivariate analysis.
(Transverse momentum conservation and rotational symmetry can reduce
the number of degrees of freedom by 3, but it does not hurt to include some redundant information.)
More precisely, we input the $(p_T, \eta, \phi)$ of each object at a
Boosted Decision Tree, which is easy to do with
TMVA~\cite{tmva} package for ROOT~\cite{root}.
The results can be taken as a
best case, to which our single variable cuts can be compared.
(To be honest, we arrived at this single variable partly by
observing which variables TMVA found most important).

\begin{figure}
\begin{center}
\begin{tabular}{cc}
\includegraphics[width=0.48\textwidth]{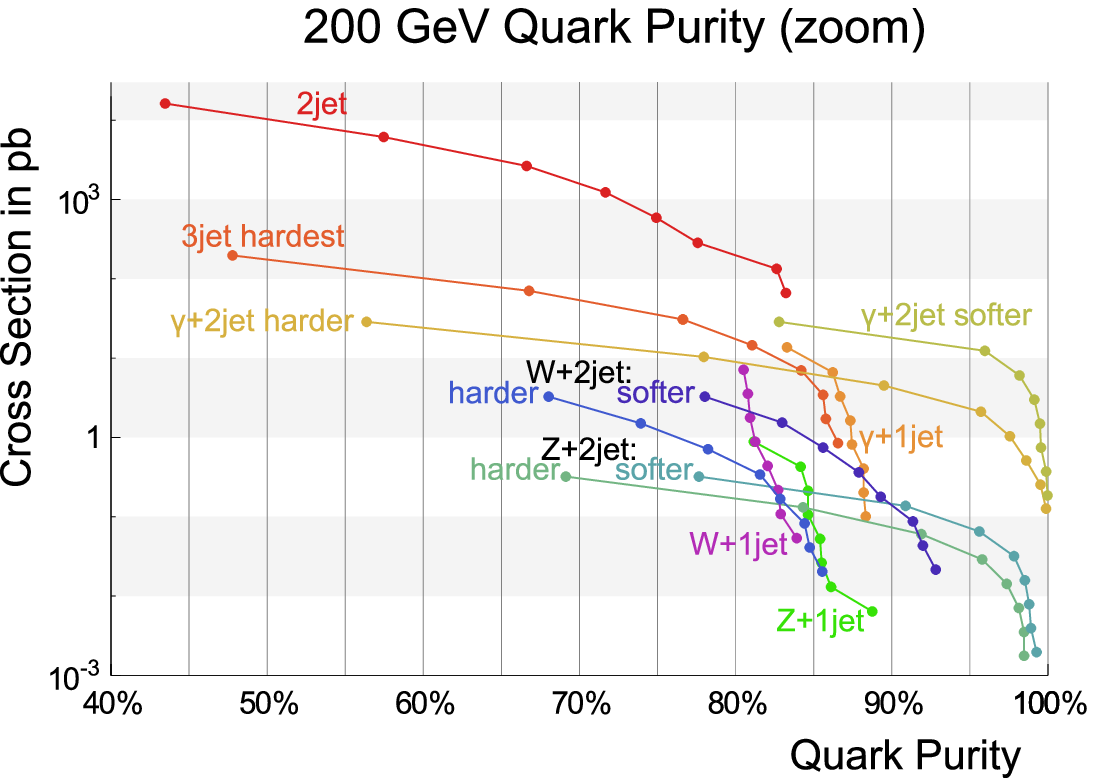}  &
\includegraphics[width=0.48\textwidth]{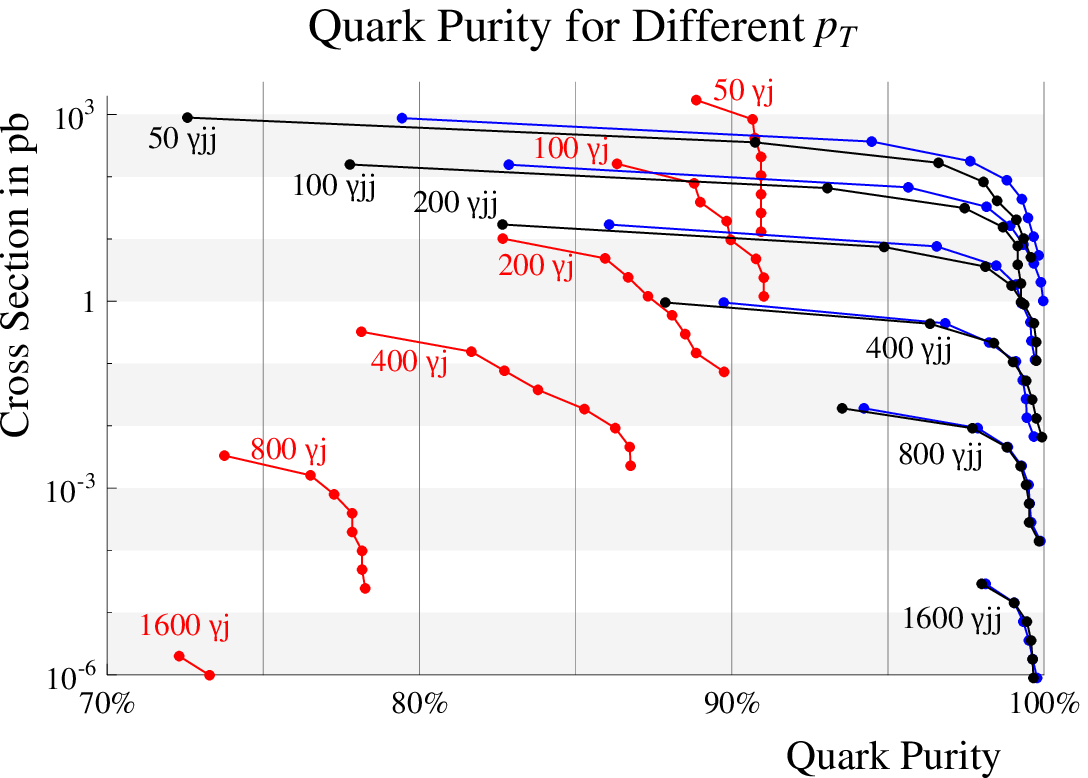}
\end{tabular}
\caption{
Cross section as a function of quark purity.
The \textbf{left} panel shows the purity for the different samples with a 200\,GeV cut on all non-$b$ jets.
The different points correspond to different cuts placed on a Boosted Decision Tree output,
trained to optimize the quark purity.
The leftmost dots of each sample are the uncut purities, and each successive dot
corresponds to cutting the number of events in half.
By the final dot, which keeps 1/128$^\mathrm{th}$ of
the signal, cutting harder no longer increases the purity.
The \textbf{right} panel shows the purities for
the $\gamma$+1jet (red) and $\gamma$+2jet (blue) samples
for various $p_T$'s, where the cuts are with BDTs trained on 6 and 9 kinematic variables, respectively.
The black curves correspond to purities obtained after cutting on the
single variable
$\eta_\gamma \eta_{j_1} + \Delta R_{\gamma j_2}$.
The blue curve takes the jet closest to the photon as a starting point,
whereas the black curve takes the softer of the two jets as its starting point.
This is the reason for the lower initial purity but the same cross section.
(It was easier to find a single variable using the softer jet
rather than the jet closer to the photon.)
}
\label{fig:quark_purity_vs_xsec_Q}
\end{center}
\end{figure}

The results of the multivariate analysis for quark jet
purification are shown in
Figure~\ref{fig:quark_purity_vs_xsec_Q}. On the left side is
the results for 200\,GeV jets, cutting on the BDT output.
Note that, as anticipated,
the $\gamma$+1jet cannot be purified much --- putting harsher cuts
hits a wall and eventually just kills the cross section. On the
right, we focus on just the $\gamma$+1jet and $\gamma$+2jet
samples for all $p_T$. The red curves are the BDT output using 6 inputs for
$\gamma$+1jet, the blue curves BDT with 9 inputs for
$\gamma$+2jets, and the black curves for our single variable
$\eta_\gamma \eta_{j 1} + \Delta R_{\gamma j_2}$. It
is nice that the single variable does as well as the
comprehensive analysis using the 9 BDT inputs.

We conclude that the best way to get a clean quark sample at low $p_T$ is to
use $\gamma$+1jet, for simplicity, or $\gamma$+2jets
at moderate to large $p_T$, cutting on the single variable
$\eta_\gamma \eta_{j 1} + \Delta R_{\gamma j_2}$. Depending
on how much cross section you are willing to sacrifice, for
200\,GeV jets, you can get 95\% quark purity at 2\,pb or 99\% purity
at 500\,nb.

\subsection{Gluon jet purification}
Next, we turn to the more difficult case of gluon jet
purification. It is more difficult because there is no starting
sample with purity above 80\%, and because there are no simple
physically motivated handles for purification. Indeed, for the
quark, we used the fact that there is a collinear $q\gamma$
singularity but no $g \gamma$ singularity to inspire a $\Delta
R_{j\gamma}$ cut. But for a gluon we cannot use the $g q$
singularity since we are trying to avoid $q$ jets all together.
The exception is samples with jets and $b$'s, where we can use
$b$-tagging information to help purify the sample. This will in
fact be relevant, but we will find that the 3- and 4-jet
samples actually work quite well, and avoid having to deal with
$b$-tagging.

\begin{figure}
\begin{center}
\includegraphics[width=0.67\textwidth]{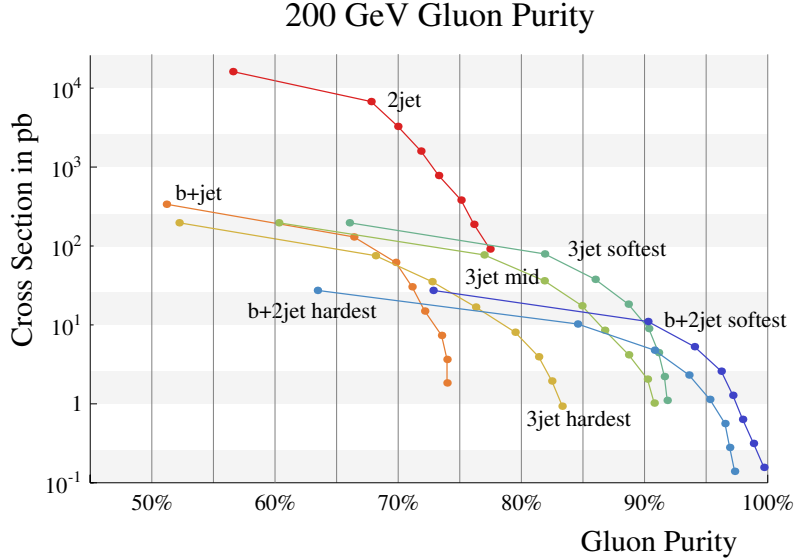}
\caption{
Cross section as a function of gluon purity
for the different samples with a 200\,GeV cut on all non-$b$ jets. The
different points correspond to different cuts placed on a
Boosted Decision Tree trained to optimize the gluon purity.
The leftmost dots of each sample are the uncut purities. There are 3 curves for
the 3-jet samples, and two for the $b$+2jet samples,
corresponding to which of the jets (from hardest to softest) is
being considered. Note the three 3-jet samples start with
identical cross sections, but higher purities are achievable
for the softer jets.
} \label{fig:gluon_purity_vs_xsec_G0}
\end{center}
\end{figure}

To begin, we start with a multivariate BDT analysis using as
inputs the ($p_T$, $\eta, \phi$) of all final
state particles. The results for the different 200\,GeV samples are shown
in Figure~\ref{fig:gluon_purity_vs_xsec_G0}. We can see that
while the $b$+2jets has good efficiency, it also has a cross
section orders of magnitude smaller than the 2-jet sample. The
3-jet sample is somewhere in between, with efficiencies
about 80\% for a cross section of 100\,pb. We will consider
these three samples in the following, as there may be
situations when each one is advantageous.

\begin{figure}
\begin{center}
\includegraphics[width=0.47\textwidth]{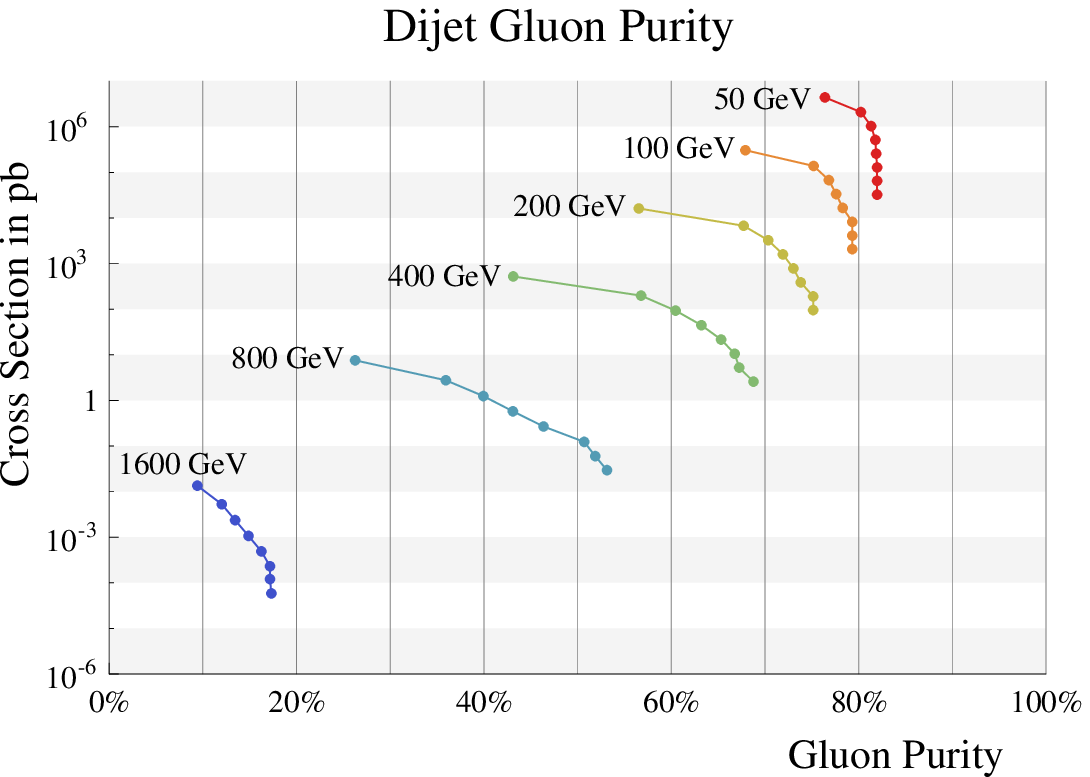}
\includegraphics[width=0.47\textwidth]{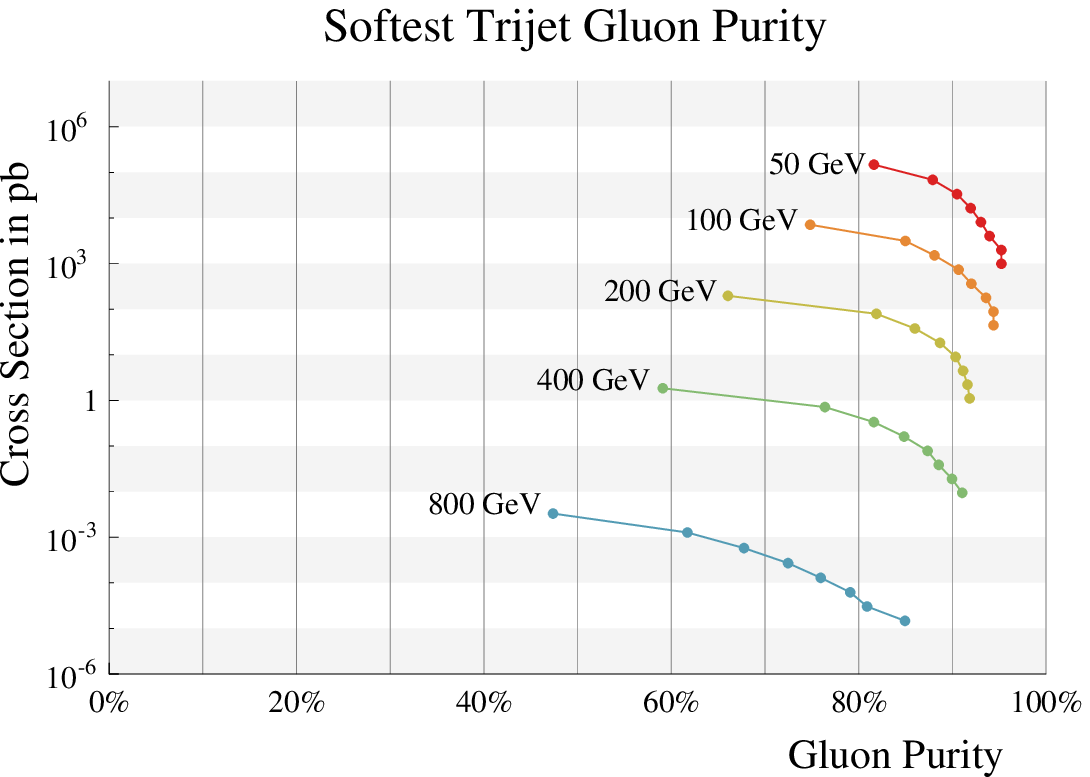}
\caption{
Gluon purities for the dijet and trijet samples for different $p_T$'s.
For each $p_T$ sample, the first dot on the top-left represents the starting
purity and cross section with no kinematic cuts.
}
\label{fig:gluon_purity_vs_xsec_G1}
\end{center}
\end{figure}

First, consider the $b$+2jet sample. Looking back at
Figure~\ref{fig:2jet_fractions}, we see that there is a
contribution from both `GG' (with $ggb$ final states) and `QG' (with $qgb$ final
states). The $ggb$ section obviously has perfect gluon
efficiency regardless of cuts. The main parton level process
contribution in the $qgb$ channel is $u b \rightarrow u b g$,
which looks like final state gluon radiation from $t$-channel
$ub\to ub$. Since we put a harder cut on the $u$ and
$g$ than the $b$, the kinematics will mostly have the $u$ going
back-to-back with the $gb$, and so the $g$ will be somewhat
softer. This explains why the starting efficiencies for the
softer jet at $p_T$=200\,GeV are around 73\%, versus 63\% for the
harder jet, as shown in see Figure~\ref{fig:gluon_purity_vs_xsec_G0}.

\begin{figure}
\begin{center}
\psfrag{Likelihood}{\quad Likelihood}
\psfrag{j2eta}{\quad  $\eta_{j_3}$}
\psfrag{j2centralj0j1deta}{\quad   $|\eta_{j_3}| - |\eta_{j_1}-\eta_{j_2}|$}
\psfrag{abseta}{\scriptsize \!\!\!\!\!\! $|\eta_{j_3}|$}
\psfrag{absj2eta}{\scriptsize \!\!\! $|\eta_{j_3}|$}
\psfrag{absj0etaMj1eta}{\scriptsize \!\!\!\!\!\!\!\!\!\!\!\!\! $| \eta_{j_1} - \eta_{j_2} |$}
\includegraphics[width=0.34\textwidth]{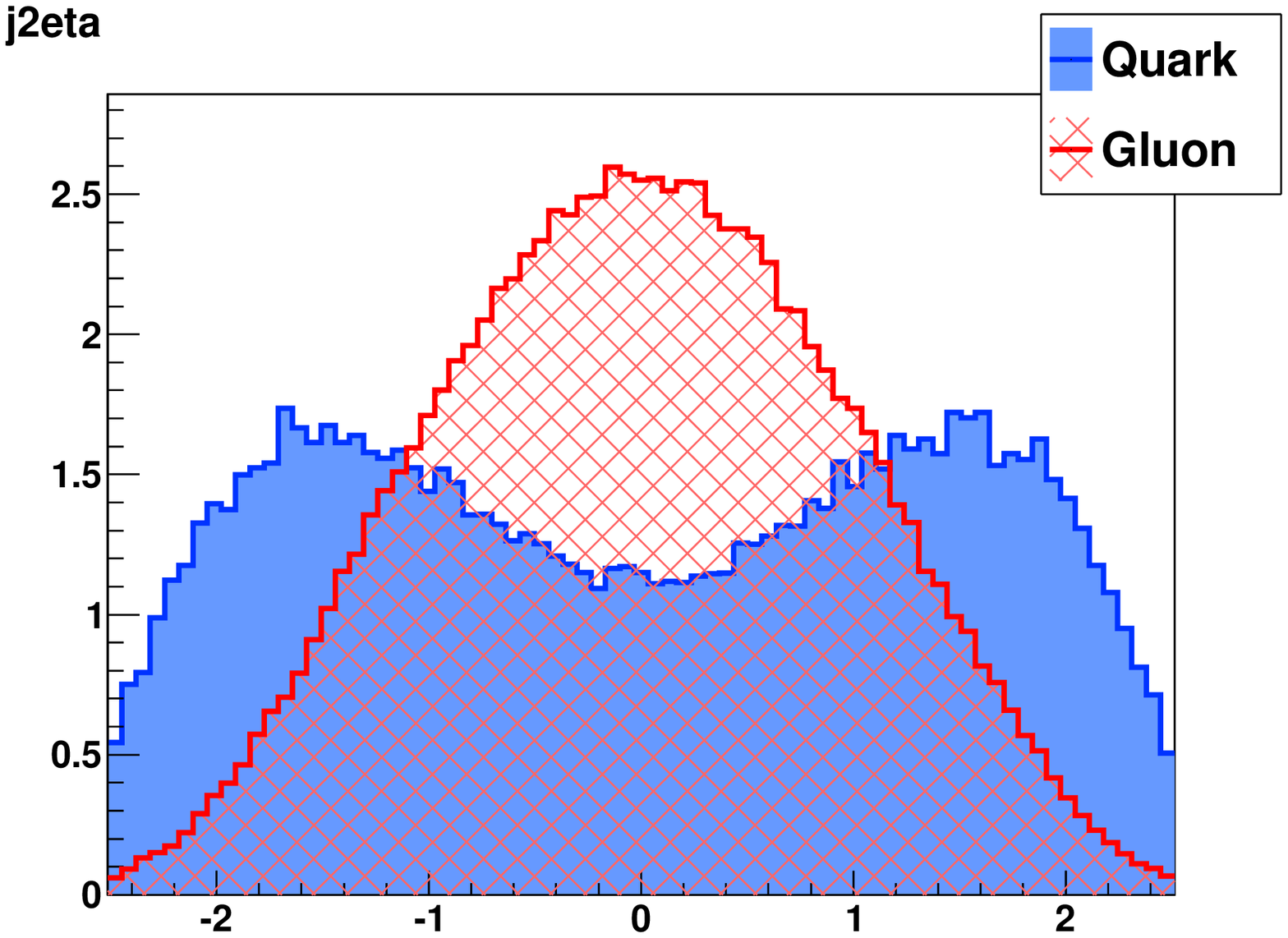}
\includegraphics[width=0.30\textwidth]{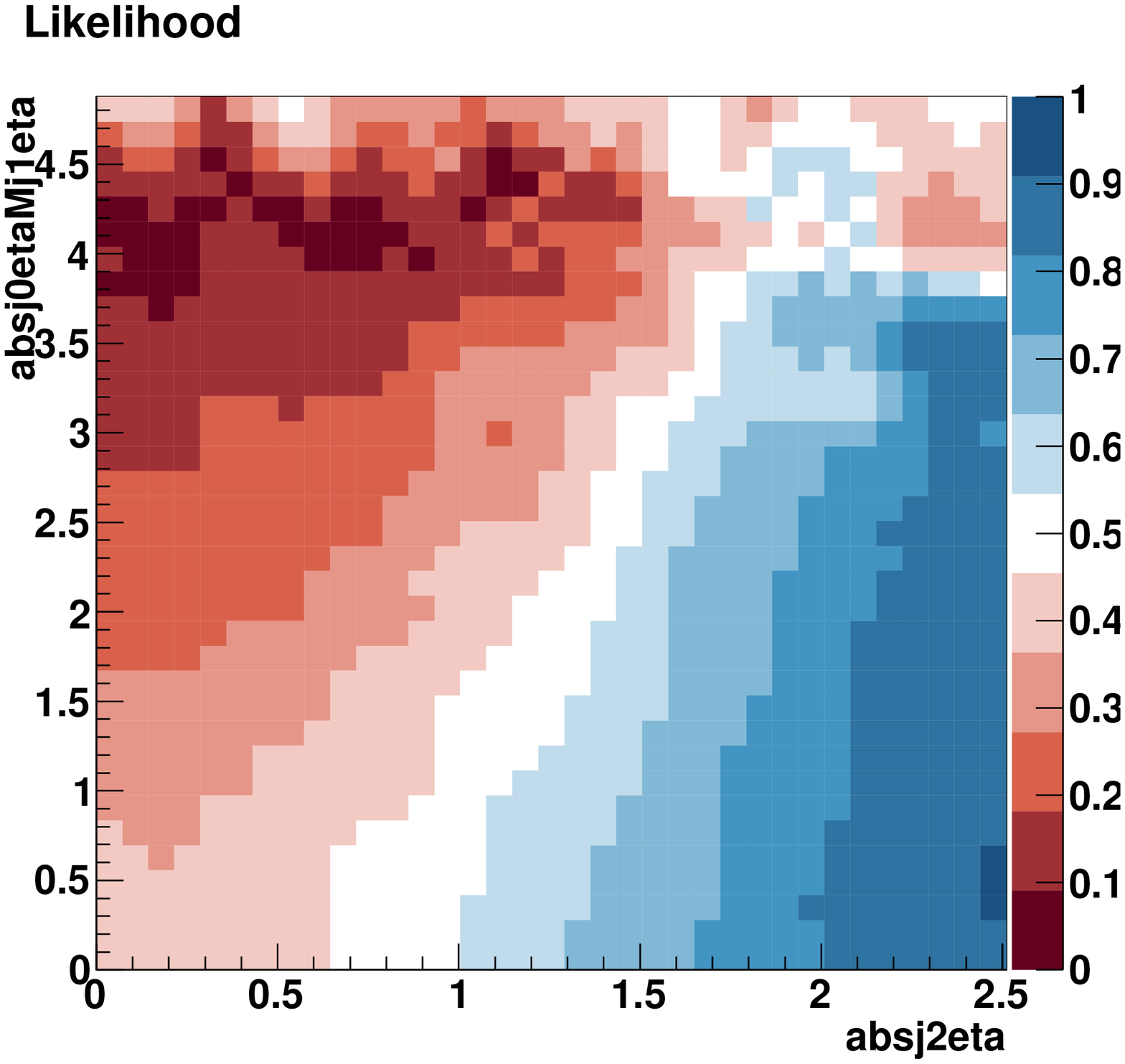}
\includegraphics[width=0.34\textwidth]{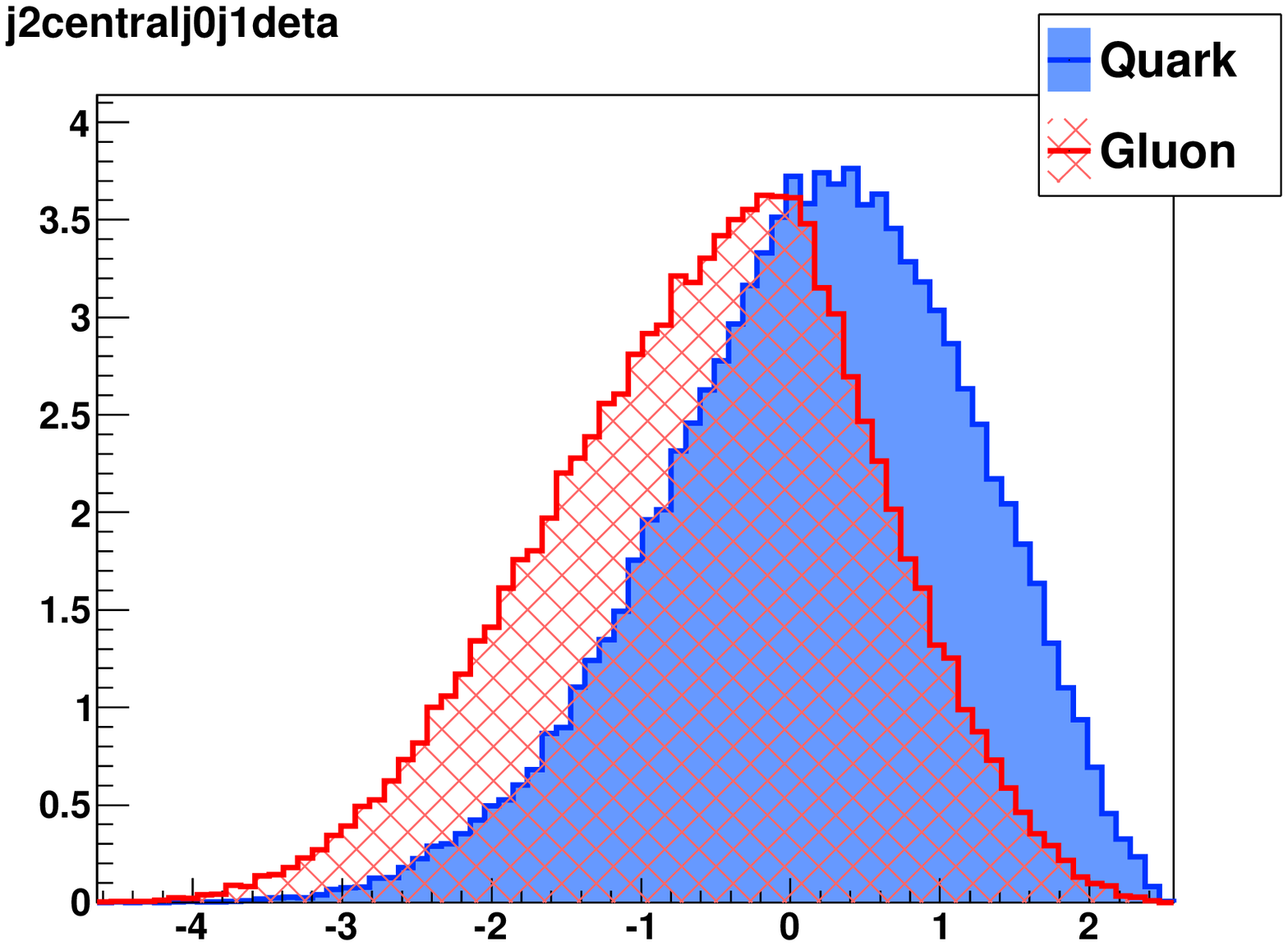}
\caption{
To purify gluons in the 3-jet sample, we look at the softest jet, which tends to be
central. It's $\eta$ is shown on the \textbf{left}.
An even better discriminant takes into account the separation of the harder two jets
and the correlation between this separation and the softest jet's $\eta$ is
shown in the \textbf{center}.
A good \emph{single} variable capturing the likelihood contours
is $|\eta_{j_3}| - |\eta_{j_1}-\eta_{j_2}|$ whose distribution is shown on the \textbf{right}.
(200\,GeV sample shown)
} \label{fig:gluon_kinematics}
\end{center}
\end{figure}

\begin{figure}[h]
\begin{center}
\includegraphics[width=0.67\textwidth]{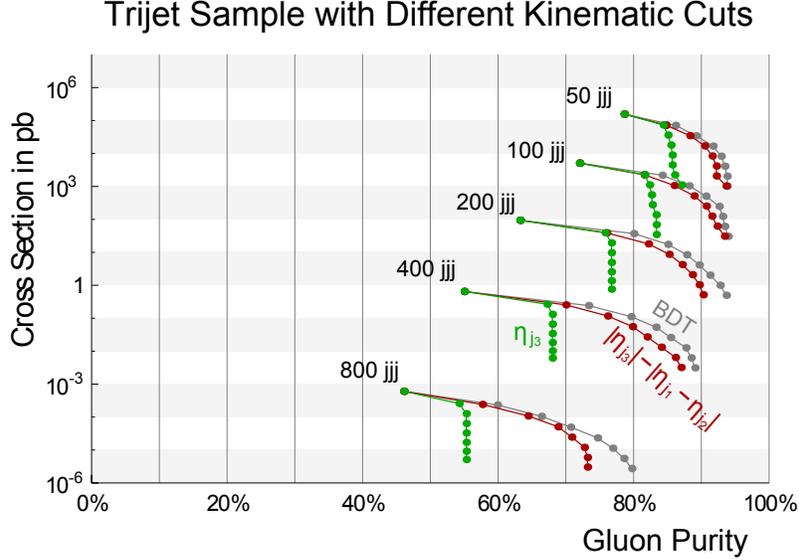}
\caption{
Cross section as a function of gluon purity.
The gray curve shows how pure the sample could be made using all the kinematic information,
through a Boosted Decision Tree.
The green curves show the result of cutting on the rapidity of the softest jet
$\eta_{j_3}$. The red curve shows that by cutting on the single variable $|\eta_{j_3}| -
|\eta_{j_1}-\eta_{j_2}|$, nearly optimal purities can be achieved, matching the BDT.
Note, all three curves agree at
their left-most points, where no cut is applied.
}  \label{fig:gluon_purity_vs_xsec_G2_3}
\end{center}
\end{figure}

\begin{figure}[h]
\begin{center}
\includegraphics[width=0.67\textwidth]{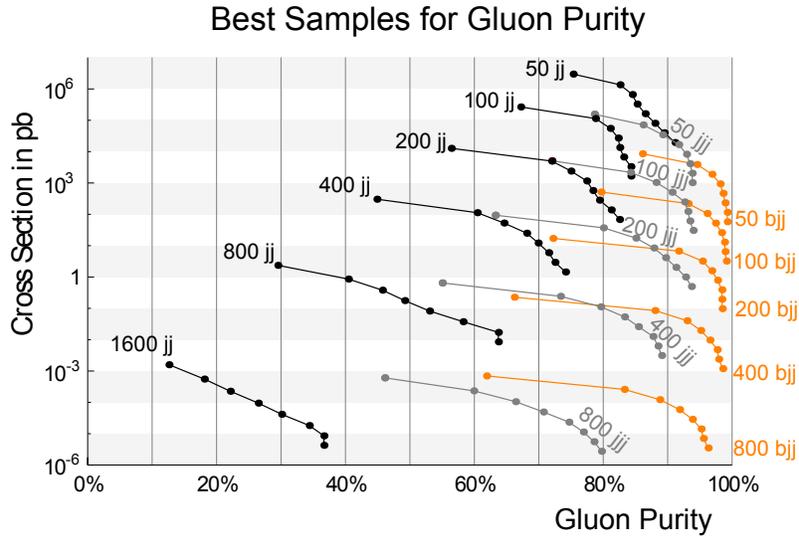}
\caption{
Cross section as a function of gluon purity. The gluon tagging
efficiencies for the dijets (black), trijets (gray), and
$b$+2jets (orange) are shown. All curves correspond to the
result of an optimal purification using a multivariate analysis.
Nearly optimal results can be reproduced in the trijet variable with a simple
cut on a single kinematical variable, as described above.
}  \label{fig:gluon_purity_vs_xsec_G2}
\end{center}
\end{figure}

The main complication in the $b$+jets samples is efficient
$b$-tagging. So far, we have assumed  perfect $b$-tagging, so
that both jets are effectively anti-$b$-tagged. In reality,
$b$-tagging can be made very tight, keeping only jets that
really look like $b$-jets or really look like non-$b$-jets.
A very tight $b$-tag will lower the cross section without
affecting the purities shown.
If looser $b$-tagging is used, the cross section will be higher
but mistags of $jjj$ and mis-anti-tags of $bbj$ make the
analysis more complicated. Note, however, that the dominant
background to $b$-jets are charm jets and from the point
of view of finding gluon jets, it is ok to treat charm jets as
$b$-jets. In many ways $b$-jets
act like gluon jets rather than like light quark jets.
For example, the OPAL experiment at LEP~\cite{Biebel:1996mc}
found $b$-jets to have more charged particles over a wider area
than light quark jets, making them similar to gluon jets in
this regard. It is therefore very important to have tight
anti-$b$-tagging on any jet used in further analysis, no matter
which starting sample it came from.
Since $b$-tagging is very detector and $p_T$
dependent, we do not attempt to include it in any quantitative
way in this tree-level study.

Next, consider the dijet and trijet samples. There is actually
a fairly strong $p_T$ dependence in the gluon fractions, as can
be seen in Figure~\ref{fig:multijet_fractions}.
As before, we begin by using full kinematic information in
Boosted Decision Trees. The result is shown in Figure~\ref{fig:gluon_purity_vs_xsec_G1}.
We see that dijets have a higher cross section, but cannot be purified
beyond a limiting value.  The trijet sample can be purified
more, but has a lower cross section since its softest jet must
be above the indicated $p_T$. While the efficiencies are not as
high as in $b$+2jets, the trijet sample can provide 90\% gluon
purity with large cross sections and few
$b$-tagging worries. A similar analysis can simplify the
kinematic cuts to a few variables.

The best single simple variable to cut on for the
softest jet in the trijet sample is the rapidity of that
jet, $\eta_{j3}$. Its distribution is shown in the left panel of of
Figure~\ref{fig:gluon_kinematics}, where we can see that the
softest jet tends to be central when it is a gluon and more
forward when it is a quark. Unfortunately just cutting on the rapidity of
the softest jet can only do so well in purifying the sample. This can be
seen from the distributions -- there is no region which is pure gluon.
To be more quantitative,
the effect of cutting on $\eta_{j3}$ is shown in
Figure~\ref{fig:gluon_purity_vs_xsec_G2_3}. The green,
representing cuts on $\eta_{j3}$ hits a hard wall for each $p_T$.

To progress further, we observe that $\eta_{j3}$ is only weakly
correlated with the rapidity difference of the other
two jets, $|\eta_{j2}-\eta_{j1}|$. The 2D distribution and the
likelihood contours are shown in the center of Figure~\ref{fig:gluon_kinematics}.
These contours are well mapped by
$|\eta_{j3}|-|\eta_{j2}-\eta_{j1}|$, which we take as our best
composite variable. Its distribution is shown on the right of this figure.
Note that, in contrast to $\eta_{j3}$, the distribution
of this composite variable has a gluon tail toward negative values. Thus, it
should be possible to put very hard cuts on it to improve efficiency.
The result is shown and contrasted to the full BDT and $\eta_{j3}$ results
in Figure~\ref{fig:gluon_purity_vs_xsec_G2_3}. We
see that cutting on this variable does nearly as well as
using the full kinematic information.

The results for the dijet, trijet and $b$+2jet samples are
summarized in Figure~\ref{fig:gluon_purity_vs_xsec_G2}. To get
very high $\sim$99\% gluon efficiencies, one needs the $b$+2jet
samples with excellent $b$-tagging.
But at 80\% or 90\%, one can instead use
trijets cutting on the discriminant
$|\eta_{j3}|-|\eta_{j2}-\eta_{j1}|$. The trijet sample has a much
larger cross section than $b$+2jets for the lower jet $p_T$ samples.

\section{Defining quark and gluon jets in QCD \label{sec:theory}}
In this section, we discuss what exactly is meant by quark and gluon jets.
%
%
We begin by considering particle decays, since they provide a context in which the concept of quark and gluon jets is more intuitive. We
then discuss how soft and collinear radiation preserves the identity of a jet as quark or gluon, and how
quark and gluon cross sections can be defined beyond leading order.

Consider a $Z$ boson which decays to 2 jets. In the limit that the jets are highly collimated and
well separated, these jets are 100\% quark jets. This is not to say that there are no gluons represented in the
jets --- beyond leading order in perturbation theory there will be many gluons, and these gluons
can have as much energy, or more, than the quarks --- but the {\it jets} coming from the $Z$-decay are still quark
jets, by definition. (There is actually zero probability for the jets to be gluon jets in this case due to Yang's theorem.)
One could also imagine a particle which would decay only to gluon jets, for example, a light Higgs boson that only couples directly to the top (the decay
would be through a top-loop). Here, the jets would unambiguously be 100\% gluon jets.
If a particle decays to 3 jets, one can ask about the quark and gluon content of the third jet as well. This
would also be well-defined to the extent that the jets are collimated and separated, which is the same extent that the jets
are representative of the hard interaction at all.
In a multiparticle cascade decay with many jets, such as in supersymmetry, one can also ask unambiguously about
the quark or gluon jet content of the various jets produced. In fact, even in QCD processes, such as $pp \to $dijets the concept
of quark and gluon jets is no more ambiguous than in decays, one is just less used to thinking about quark and gluon fractions.

When jets are highly collimated and well separated, their cross sections factorize into
the production process, for which there is no mixing between quarks and gluons, and the fragmentation
process, whereby those quark and gluon jets shower and hadronize into observable particles.
Although exact factorization proofs are not available for anything but the simplest process (Drell-Yan),
scaling arguments suggest that any violations to factorization should be negligible. Thus, the
concept of quark and gluon jets is a well-defined theoretical concept up to power corrections that scale
as $\Lambda_{\mathrm{QCD}}/E$ and $R\sim m/E$, where $R$ is the size of the jet, $E$ its energy and $m$ its mass.

As mentioned in the introduction, there is no ambiguity at leading order
in defining the fraction of quark and gluon jets in any exclusive sample. To be precise, leading order here
means the Born level, the lowest order in perturbation theory which produces the required number of jets. To be concrete,
consider for example the direct production of a hard photon, say with $p_T > 200$ GeV.
At leading order, there are two Feynman diagrams, the Compton channel: $q g \to q \gamma$ and the annihilation
channel $q\bar{q} \to g \gamma$. The ratio of the cross sections for these channels, at leading order,
tells us that 85\% of the jets produced in association with a photon will be quark jets. For more complicated
processes there is also no ambiguity as long as we are specific about which jet we mean, in an infrared safe way.
For example, we can ask about the 2nd hardest anti-$k_T$ $R=0.4$ jet in $W$+jets events. Here, the Born level is $W$+2 jets,
and the cross section ratio can be computed unambiguously (up to scale uncertainties) at leading order.

At next-to-leading order, there are virtual and real contributions. Both of these are infrared divergent and some
part of the real contributions must be added to the virtual to get a finite answer. The virtual graphs
have the same number of jets as the Born level, and so whether they contribute to the quark or gluon jet cross section
is similarly unambiguous. The real graphs can be split into a contribution containing the infrared divergent regions
and a hard remainder. The infrared divergences are soft or collinear, and in either limit the identity of the jet
as quark or gluon is conserved. In the soft limit, the interactions of gluons are Eikonal and factorize
off, again leaving the quark or gluon nature of the jet unchanged.
In the collinear limit, helicity is conserved. So one can treat the helicity of a jet as a conserved quantum number
which is necessarily different for quark and gluon jets. Moreover, for any infrared-safe jet definition, a collinear gluon
emitted in the singular region must go into the jet,
so the overall baryon number of the jet (number of quarks minus number of antiquarks) is conserved. Hard emissions
must produce another jet, at least in the approximation where the jets are highly collimated, which is where factorization
holds.\footnote{There may be additional ``non-global'' contributions, from configurations where a hard gluon splits into two quarks
and one of those ends up a jet. Whether non-global logs are relevant or not is a question about the observable, such
as the jet mass, not about whether the jets are quark or gluon. Quark or gluon jets are defined to the extent that
factorization holds, and non-global logs would violate factorization.
More information on non-global logs can be found in ~\cite{Dasgupta:2001sh,Kelley:2011ng,Hornig:2011iu}.}
So the infrared-singular parts of the real emission contributions do not change whether the jet is quark or gluon
and therefore the quark or gluon fraction can be defined at higher orders in perturbation theory.

To all orders in perturbation theory, the factorization into quark and gluon production can be simplified by the
use of operators in Soft-Collinear Effective Theory~\cite{Bauer:2000yr,Bauer:2001yt}.
For example, for direct photon production~\cite{Becher:2009th}, there are
6 production channels, with initial states $q q, \qbar \qbar, q \qbar, q g, g g$ and $\qbar g$. Each channel
has two spin structures, corresponding to the cases when the quarks have equal or opposite spin.
For example, in the $q\qbar \to g \gamma$ channel,
the operators are
\begin{equation}
 {\mathcal O}^{S\nu}_{q \qbar} = \bar{\chi}_2 {\mathcal A}_\perp^\nu \chi_1,
\qquad
 {\mathcal O}^{T\nu}_{q \qbar} = \bar{\chi}_2 \sigma_{\mu \nu} {\mathcal A}_\perp^\mu \chi_1, \label{SCETops}
\end{equation}
 So there are 12 operators total relevant for matching at the Born level. The fields
$\chi$ and ${\mathcal A}$ are collinear quarks and gluons with associated collinear Wilson lines.
For simplicity, these are called jet fields. More details can
of the notation can be found in~\cite{Becher:2009th}.

The point of the SCET notation is that it gives a precise definition to what we have been calling quark and gluon jets.
It therefore lets us define the quark and gluon jet fractions exactly, as ratios of matrix elements of operators
with quark or gluon jet fields.
In the limit where factorization holds, there is no mixing between operators with different jet fields, or even of
fields with different spin. For example, in direct photon, when the photon is very energetic there
is only phase space for it to recoil against a single jet.
In this limit, the process is exactly described by the operators in Eq.~\eqref{SCETops} and the other 10 operators for the other channels.
The mixing between the operators is power suppressed. To add some concreteness to the discussion,
at leading order, the jet recoiling against at 300\,GeV photon is 82.3\% quark. At NLO, it is 84.6\% quark and at NNLO 85.1\% quark.
The leading order prediction is a very good approximation to more precise values, since the radiative corrections largely drop out of
the fraction.

In summary, in this section we have explained how the quark and gluon jet fraction is exactly defined in a limit in which
the production of jets factorizes into an incoherent sum of different channels. This gives precisely calculable
cross sections, and hence a well-defined quark-to-gluon jet fraction.
%

\section{Conclusions \label{sec:conclusion}}
In this paper, we have systematically explored which processes at a
proton collider can be exploited to give
pure samples of quark and gluon jets.
We found that a 98\% pure quark jet sample is achievable by starting with the
softer jet in $\gamma$+2jets and cutting on the combined
kinematic variable $\eta_\gamma \eta_{j_1} + \Delta R_{\gamma j_2}$. The corresponding cross sections are
around 10\,pb for $p_T \ge 100$\,GeV, 1\,pb for $p_T \ge 200$\,GeV, or
0.1\,pb for $p_T \ge 400$\,GeV quark jets. More quark purity information is in
Figure~\ref{fig:quark_purity_vs_xsec_Q}.

Gluon jets are more difficult to purify. We found that the $b$+2jets sample
provides the best results under ideal conditions.
Unfortunately, to get such pure gluon jet samples requires
a excellent $b$-tagger, and a realistic analysis can only be
done with details of the particular experiment
and $b$-tagging method. The next best thing, is to use
the softest jet in 3jet events.
This has a higher cross section than the $b$+2jets sample, but cannot
achieve quite as high purities.
Cutting on the
combined variable $|\eta_{j3}|-|\eta_{j2}-\eta_{j1}|$, the trijet sample
can provide 100\,pb at 93\% purity for 100\,GeV gluon jets, 1\,pb
for 90\% purity 200\,GeV jets, or 10\,fb of 85\% purity
400\,GeV jets.  More gluon purity information is in
Figure~\ref{fig:gluon_purity_vs_xsec_G2}.

The fraction of quark and gluon jets, which we have calculated in this paper at leading order in perturbation theory, is well-defined
theoretical concept, up to power corrections in the jet size. These power corrections are suppressed when the jets are hard and well-separated.
The quark-to-gluon jet fraction is a theoretical concept, not directly observable, but it is an extremely useful theoretical concept.
The observables are the jet properties in a given sample, which correlate with the quark or gluon
jet fraction. These properties, such as mass of the hardest jet, can in principle also be calculated.
Certain regions of phase space, the ones with pure samples of quark or gluon jets discussed in this paper, should allow us to test calculations and calibrate simulations
of jet properties more efficiently.
With the better experimental handle on jet properties arising from the study of these samples, we will be better prepared to extract properties
of fundamental standard-model or beyond-the-standard-model physics encoded in hadronic events.
%
%


\section*{Acknowledgments}
We thank Salvatore Rappoccio and Michael Kagan for discussions
about CMS and ATLAS, and David Krohn and Peter Wittich for comments on the manuscript.
JG thanks Michigan State for its speaking
invitation and ensuing discussions that inspired this work,
along with a night stuck in the Detroit airport generating samples.
This work was supported in part by the Department of Energy
under grant DE-SC003916. Computations for this paper were
performed on the Odyssey cluster supported by the FAS Research
Computing Group at Harvard University.

\end{document}